\documentclass[12pt,letterpaper]{article}

\usepackage{xpatch}
\makeatletter
\xpatchcmd{\ps@headings}{\sectionmarkformat\fi}{\sectionmarkformat}{}{} 

\usepackage{amssymb}
\usepackage{enumitem}
\usepackage{mathptmx}
\usepackage{color}
\usepackage[margin=1in]{geometry}
\usepackage{amsmath}
\usepackage{hyperref}
\usepackage{wrapfig}
\usepackage{graphicx}
\usepackage{upgreek}
\usepackage{textcomp}
\usepackage[utf8]{inputenc}
\usepackage[font=small,labelfont=bf]{caption}[2003/12/20]
\usepackage{sectsty}
\usepackage{lineno}
\usepackage[compact]{titlesec}
\usepackage{xcolor}
\usepackage{sidecap}
\usepackage[strict]{chngpage}
\usepackage[sort&compress,numbers]{natbib}
\usepackage{authblk}
\usepackage{abstract}

\allsectionsfont{\normalfont\sffamily\bfseries}

\title{Cosmic-ray Antinuclei as Messengers of New Physics: Status and Outlook for the New Decade}

\author[a]{P. von Doetinchem\footnote{Corresponding authors: philipvd@hawaii.edu, kmperez@mit.edu.}}
\affil[a]{Department of Physics and Astronomy, University of Hawaii at Manoa, 2505 Correa Rd, Honolulu, HI 96822} 

\author[*b]{K. Perez}
\affil[b]{Department of Physics, Massachusetts Institute of Technology, 77 Massachusetts Ave, Cambridge, MA 02139, USA} 

\author[c]{T. Aramaki}
\affil[c]{Stanford Linear Accelerator Center, 2575 Sand Hill Rd, Menlo Park, CA 94025, USA} 

\author[d]{S. Baker} 	
\affil[d]{Imperial College London, London, SW7 2AZ, UK} 

\author[e]{S. Barwick}
\affil[e]{Department of Physics \& Astronomy, University of California at Irvine, 4129 Frederick Reines Hall, Irvine, CA 92697, USA} 

\author[f]{R. Bird}
\affil[f]{Department of Physics and Astronomy , University of California at Los Angeles, 475 Portola Plaza, Los Angeles, CA 90095, USA} 

\author[g]{M. Boezio}	
\affil[g]{INFN, Sezione di Trieste, Padriciano 99, 34149 Trieste, Italy}

\author[h]{S.E. Boggs}
\affil[h]{Department of Physics, University of California at San Diego, 9500 Gilman Dr., La Jolla, CA 90037, USA}

\author[i]{M. Cui}
\affil[i]{Purple Mountain Observatory, Yuanhua Road, Qixia District, Nanjing 210033, China}

\author[a]{A. Datta}

\author[j,k]{F. Donato}	
\affil[j]{Department of Physics, University of Turin, Via Pietro Giuria, 1, 10125 Torino, Italy}
\affil[k]{INFN, Sezione di Torino, Via Pietro Giuria, 1, 10125 Torino, Italy}

\author[l,m]{C. Evoli}
\affil[l]{Gran Sasso Science Institute, Viale Francesco Crispi 7, 67100 L'Aquila, Italy}
\affil[m]{INFN, Laboratori Nazionali del Gran Sasso, Via G. Acitelli, 22, 67100 Assergi, L'Aquila, Italy}

\author[n]{L. Fabris}
\affil[n]{Isotope and Fuel Cycle and Technology Division, Oak Ridge National Laboratory, PO BOX 2008, Oak Ridge, TN 37831, USA}

\author[o]{L. Fabbietti}
\affil[o]{Department of Physics, Technical University of Munich, James-Franck Str. 1, 85748 Garching, Germany}

\author[p]{E. Ferronato Bueno}
\affil[p]{Kapteyn Astronomical Institute, Rijksuniversiteit Groningen, Landleven 12, 9717 AD Groningen, The Netherlands}

\author[j,k]{N. Fornengo}

\author[q]{H. Fuke}
\affil[q]{Institute of Space and Astronautical Science, Japan Aerospace Exploration Agency (ISAS/JAXA), Sagamihara, Kanagawa 252-5210, Japan}

\author[a]{C. Gerrity}

\author[a,r]{D. Gomez Coral}
\affil[r]{Institute of Physics, National Autonomous University of Mexico, Circuito de la investigación científica, C.U. 04510, Ciudad de México, Mexico.}

\author[s]{C. Hailey}
\affil[s]{Department of Physics, Columbia University, 500 W 120th St, New York, NY 10027, USA}

\author[t,u,v]{D. Hooper}
\affil[t]{Theoretical Astrophysics, Fermi National Accelerator Laboratory, Wilson and Kirk Rds, Batavia, IL 60510, USA}
\affil[u]{Department of Astronomy and Astrophysics, University of Chicago, 5640 S. Ellis Ave, Chicago, IL 60637, USA}
\affil[v]{Kavli Institute for Cosmological Physics, University of Chicago, 5640 S. Ellis Ave, Chicago, IL 60637, USA}

\author[w]{M. Kachelriess}
\affil[w]{Department of Physics, Norwegian University of Science and Technology, 7491 Trondheim, Norway }

\author[j,k,x]{M. Korsmeier}
\affil[x]{Institute for Theoretical Particle Physics and Cosmology, RWTH Aachen University, 52056 Aachen, Germany}

\author[q]{M. Kozai}

\author[g,y]{R. Lea}
\affil[y]{Dipartimento di Fisica dell’Università Trieste, Via Valerio 2, 34127 Trieste, Italy}

\author[z,A]{N. Li}
\affil[z]{CAS Key Laboratory of Theoretical Physics, Institute of Theoretical Physics, Chinese Academy of Sciences, Beijing 100190, China}
\affil[A]{University of Chinese Academy of Sciences, No.19A Yuquan Road, Shijingshan District, Beijing 100049, China}

\author[h]{A. Lowell}

\author[B,C]{M. Manghisoni}
\affil[B]{INFN, Sezione di Pavia, Via Agostino Bassi 6, 27100 Pavia, Italy}
\affil[C]{Dipartimento di Ingegneria Industriale, Universit\`{a} di Bergamo, Viale Marconi 5, 24044 Dalmine, Italy}

\author[D,E]{I.V. Moskalenko}
\affil[D]{Hansen Experimental Physics Laboratory, Stanford University, 452 Lomita Mall, Stanford, CA 94305, USA}
\affil[E]{Kavli Institute for Particle Astrophysics and Cosmology, Stanford University, CA 94305, USA}

\author[g]{R. Munini}

\author[a,F]{M. Naskret}
\affil[F]{Institute of Theoretical Physics, University of Wroclaw, pl. M. Borna 9, 50-204 Wroclaw, Poland}

\author[a]{T. Nelson}	

\author[G]{K.C.Y. Ng}
\affil[G]{Department of Particle Physics and Astrophysics, Weizmann Institute of Science, Rehovot 76100, Israel} 

\author[H]{F. Nozzoli}
\affil[H]{INFN, Trento Institute for Fundamental Physics and Applications, Via Sommarive, 14, 38123 Povo, Italy}

\author[I]{A. Oliva}
\affil[I]{INFN, Sezione di Bologna, Via Irnerio 46, Bologna 40126, Italy}

\author[f]{R.A. Ong}

\author[J]{G. Osteria}
\affil[J]{INFN, Sezione di Napoli, Strada Comunale Cinthia, 80126 Naples, Italy}

\author[K]{T. Pierog}
\affil[K]{Institute for Nuclear Physics, Karlsruhe Institute of Technology, Hermann-von-Helmholtz-Platz 1, 76344 Eggenstein-Leopoldshafen, Germany}

\author[L]{V. Poulin}
\affil[L]{Laboratoire Univers \& Particules de Montpellier, CNRS, Universit\'e de Montpellier, Place Eug\`ene Bataillon, 34095 Montpellier Cedex 05, France}

\author[M]{S. Profumo}
\affil[M]{Department of Physics and Santa Cruz Institute for Particle Physics, University of California, Santa Cruz, CA 95064, USA} 

\author[o]{T. Pöschl}

\author[f]{S. Quinn}

\author[B,C]{V.~Re}
 
\author[b]{F. Rogers}

\author[f]{J. Ryan}	

\author[s]{N. Saffold}

\author[N,O]{K. Sakai}	
\affil[N]{ NASA-Goddard Space Flight Center), 8800 Greenbelt Rd, Greenbelt, MD 20771, USA}
\affil[O]{CRESST, University of Maryland, Baltimore County, MD 21250, USA}

\author[P]{P. Salati}
\affil[P]{Laboratoire d'Annecy-le-Vieux de Physique Théorique, 9 Chemin de Bellevue, 74940 Annecy, France}

\author[Q]{S. Schael}
\affil[Q]{I. Physikalisches Institut, RWTH Aachen University, Sommerfeldstr. 14, 52074 Aachen, Germany} 

\author[o]{L. Serksnyte}

\author[a]{A. Shukla}

\author[a]{A. Stoessl}

\author[w]{J. Tjemsland}	

\author[R]{E. Vannuccini}
\affil[R]{INFN, Sezione di Firenze, 50019 Sesto Fiorentino, Florence, Italy} 

\author[p]{M. Vecchi}

\author[S]{M.W. Winkler}
\affil[S]{The Oskar Klein Centre for Cosmoparticle Physics, Department of Physics, Stockholm University, Alba Nova, 10691 Stockholm, Sweden} 

\author[c]{D. Wright}

\author[b]{M. Xiao}

\author[T]{W. Xu}
\affil[T]{Department of Physics, Harvard University, 17 Oxford St, Cambridge, MA, 95129, USA} 

\author[q]{T. Yoshida}

\author[g]{G.~Zampa}

\author[H,U]{P. Zuccon}
\affil[U]{Department of Physics, University of Trento, Via Sommarive 14, 38123 Povo, Italy}

\date{}

\begin{document}
\maketitle

\begin{abstract}

The precise measurement of cosmic-ray antinuclei serves as an important means for identifying the nature of dark matter and other new astrophysical phenomena, {and could be used with other cosmic-ray species to understand} cosmic-ray production and propagation in the Galaxy.  
For instance, low-energy antideuterons {would} provide a ``smoking gun" signature of dark matter annihilation or decay, essentially free of astrophysical background.
Studies in recent years have emphasized that models for cosmic-ray antideuterons must be considered together with the abundant cosmic antiprotons and any potential observation of antihelium.
Therefore, a second dedicated Antideuteron Workshop was organized at UCLA in March 2019, bringing together a community of theorists and experimentalists to review the status of current observations of cosmic-ray antinuclei, the theoretical work towards understanding these signatures, and the potential of upcoming measurements to illuminate ongoing controversies.
This review aims to synthesize this recent work and present implications for the upcoming decade of antinuclei observations and searches. 
This includes discussion of a possible dark matter signature in the AMS-02 antiproton spectrum, the most recent limits from BESS Polar-II on the cosmic antideuteron flux, and reports of candidate antihelium events by AMS-02; recent collider and cosmic-ray measurements relevant for antinuclei production models; the state of cosmic-ray transport models in light of AMS-02 and Voyager data; and the prospects for upcoming experiments, such as GAPS. 
This provides a roadmap for progress on cosmic antinuclei signatures of dark matter in the coming years.

\end{abstract}

\pagebreak

\tableofcontents

\section{Introduction}
\label{s-intro}

Cosmic-ray antinuclei, i.e. antiprotons, antideuterons, and antihelium, offer unique sensitivity to annihilating and decaying dark matter, probing a variety of dark matter models that evade or complement collider, direct, or other cosmic-ray searches.
This article reviews the status of the field of cosmic-ray antinuclei measurements, including both interpretation of current results and the potential for breakthrough in the coming decade. 
These discussions were presented at the ``$2^{\text{nd}}$ Cosmic-ray Antideuteron Workshop'' conducted in March 2019\footnote{\url{https://indico.phys.hawaii.edu/e/dbar19}}. 

Since the 1970s, antiproton experiments have developed from a first-time detection~\cite{1979PhRvL..43.1196G,1979ICRC....1..330B} into an important tool for dark matter {studies} and cosmic-ray physics.
The BESS~\cite{2013AdSpR..51..227B,besspbar}, CAPRICE98~\cite{2000ApJ...534L.177B}, PAMELA~\cite{2010PhRvL.105l1101A,Adriani:2012paa}, and AMS-02~\cite{PhysRevLett.117.091103} antiproton results have all been used to deliver leading constraints on dark matter models, as well as astrophysical production and propagation scenarios~\cite{1998ApJ...499..250S,1998A&A...338L..75M,1999ApJ...526..215B,2002ApJ...565..280M,2003ApJ...586.1050M,2006ApJ...642..902P,2012ApJ...752...68V,Fornengo:2013osa,pbarci,Hooper:2014ysa,Boudaud:2014qra,Cerdeno:2014cda,2015arXiv150405937H,2015arXiv150604145K,Chen:2015cqa,Winkler:2017xor}. 
Recently, a possible excess in the AMS-02 low-energy antiproton spectrum that is consistent with 20--80\,GeV dark matter has been uncovered~\cite{PhysRevLett.118.191101,PhysRevLett.118.191102,Reinert:2017aga,Cuoco:2019kuu,Cholis:2019ejx}.
Intriguingly, this excess is consistent with many of the same models indicated by the long-standing excess of gamma-rays from the Galactic center~\cite{Cholis:2019ejx}. 
However, the significance of this possible dark matter signal depends on the interpretation of the theoretical and experimental systematic uncertainties, in particular careful treatment of any energy correlations~\cite{Boudaud:2019efq} and background from standard astrophysical processes.

In contrast to the high-statistics antiproton measurements, searches for cosmic antideuterons are focused on a first-time discovery, offering a powerful new method of probing cosmic physics.
The unique strength of a search for antideuterons lies in their ultra-low astrophysical background, particularly at low energies.
Even taking into account the constraints from antiprotons, a variety of dark matter models predict an antideuteron flux that is orders of magnitude above the astrophysical background in the energy range below a few GeV/$n$~\cite{Donato:1999gy,Baer:2005tw,Donato:2008yx,Duperray:2005si,Ibarra:2012cc,Ibarra2013a,Fornengo:2013osa,Dal:2014nda, Korsmeier:2017xzj,Tomassetti:2017qjk,Lin:2018avl,Li:2018dxj}. This is in contrast to other cosmic-ray dark matter searches, such as those using positrons or gamma-rays, which rely on small {excesses} on top of {considerable} astrophysical backgrounds. 
{For the case of positrons an excess around 100\,GeV is visible~\cite{2019PhRvL.122d1102A}, but  the main difficulty {with the identification of a potential dark matter signal} is that there are many other {conventional} sources that can produce positrons in this energy range.}
The ultra-low astrophysical background {for dark matter searches with antideuterons} is a consequence of the kinematics of antideuteron formation. 
In standard cosmic-ray physics, antinuclei are produced when cosmic-ray protons or antiprotons interact with the interstellar {gas}~\cite{Duperray:2005si,Ibarra2013a}. The high energy threshold for antideuteron production and the steep energy spectrum of cosmic rays mean that there are {fewer} particles with sufficient energy to produce an antinucleus, and those that are produced have relatively large kinetic energy.
Therefore, low-energy antideuteron searches are virtually free from conventional astrophysical background. Meanwhile, low-energy antiproton measurements are crucial to these antideuteron searches, as any antideuteron detection must be consistent with antiproton search results.
In principle, the production of antiprotons and heavier antinuclei in collisions of cosmic rays with the interstellar medium or dark matter annihilation or decay is tightly coupled, resulting in a degeneracy of the individual sensitivities to dark matter. However, the antinuclei formation processes have large uncertainties, and thus considerable freedom for the prediction of antideuteron and antihelium nuclei fluxes exists.
The coming decade is an exciting time for experimental searches for cosmic antideuterons, {since} experiments {are} beginning to be sensitive to viable dark matter models.
AMS-02 is accumulating large statistics on the International Space Station, and the GAPS Antarctic balloon mission, the first experiment optimized specifically for low-energy ($E < 0.25$\,GeV/$n$) cosmic antideuterons, is preparing for its initial flight in late 2021.

These experiments are also sensitive to antihelium nuclei.
Recently, the AMS-02 collaboration has announced the observation of several candidate antihelium-3 and antihelium-4 events~\cite{antihe,antihe2,antihe3}. Data taking, analyses, and interpretation of these possible signals are still ongoing within the AMS-02 collaboration. {For the purpose of this review, it needs to be highlighted that the antihelium  detection has not yet been confirmed. Therefore, any discussion of cosmic-ray antihelium fluxes in this review should be seen as theoretical case study.}
Although these results are not yet published, they have prompted significant theoretical work on the implications for dark matter models and predicted antideuteron and antiproton fluxes. 
Antihelium arriving from antimatter-dominated regions of the universe is already nearly excluded~\cite{Adams:1997ym,Cohen:1997ac}.
Proposed {models} span a range including modified antihelium formation models~\cite{Blum:2017qnn,Tomassetti:2017qjk}, dark matter annihilation~\cite{Cirelli:2014qia, Carlson:2014ssa, Coogan:2017pwt, Korsmeier:2017xzj, Li:2018dxj}, or emission from a nearby antistar~\cite{2018arXiv180808961P}. 
Though these candidates are tentative, requiring verification or refutation with improved analysis from AMS-02 and a complementary experimental technique such as GAPS, a positive signal would be transformative, refashioning the field of cosmic-ray physics and potentially revolutionizing our understanding of Big Bang nucleosynthesis.

This review proceeds as follows.
Sec.~\ref{s-exp} reviews the status of current experiments sensitive to cosmic-ray antiproton, antideuteron, and antihelium nuclei. This is followed in Sec.~\ref{s-impact} by a discussion of the interpretation of and predictions based on the current data. Key ingredients for this interpretation are uncertainties in the antinuclei formation and cosmic-ray propagation models. Sec.~\ref{s-uncertain} discusses the status of theoretical and experimental strategies to understand and reduce these uncertainties. Sec.~\ref{s-future} presents an overview of new experimental ideas for the decades beyond the ongoing experimental efforts.

\section{Experimental Status\label{s-exp}}

\subsection{The BESS Experiment\label{s-bess}}

\begin{figure}
  \begin{center}
    \includegraphics[width=1\linewidth]{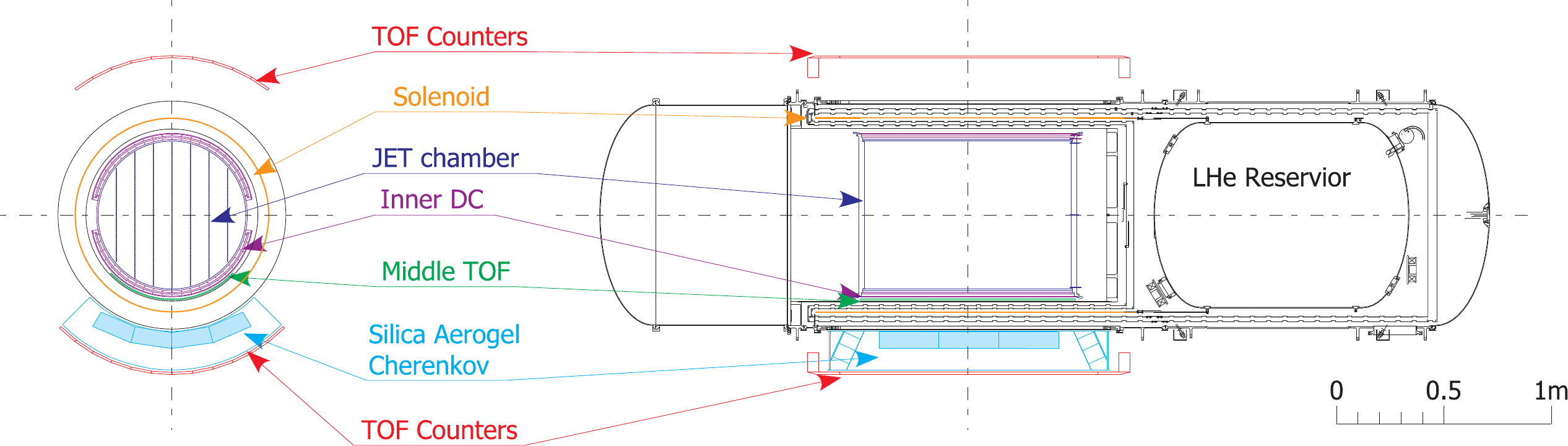}
    \end{center}
\caption{\label{f-fig1} The BESS-Polar II experiment consists of inner drift chambers and a jet-type drift tracking chamber, surrounded by a solenoid magnet, aerogel Cherenkov counter, and time-of-flight system. Reprinted with permission from~\cite{2012PhRvL.108m1301A}.}
\end{figure}

The BESS program (1993-2008)~\cite{2013AdSpR..51..227B}, culminating in the BESS-Polar instrument, exploits particle tracking in a solenoidal magnetic field to identify antimatter. 
The original BESS-Polar experiment flew over Antarctica in late 2004. The BESS-Polar II experiment collected 24.5\,days of Antarctic flight data from December 2007 to January 2008~\cite{besspbar,2012PhRvL.108m1301A}. 

BESS-Polar II, shown in Fig.~\ref{f-fig1}, consists of a 0.8\,T solenoidal magnet, filled by inner drift chambers (IDC) and a jet-type drift tracking chamber (JET), and surrounded by an aerogel Cherenkov counter (ACC) and a time-of-flight (TOF) system composed of scintillation counter hodoscopes. These components are arranged in a coaxial cylindrical geometry, providing a large geometric acceptance of 0.23\,m$^{2}$\,sr.  
Tracking is performed by fitting up to 48 hit points in the JET and 4 hit points in the IDC, resulting in a magnetic-rigidity resolution of $0.4\%$ at 1\,GV and a maximum detectable rigidity of 240\,GV. The upper and lower scintillator hodoscopes provide TOF and $\text d E/\text d x$ measurements as well as trigger signals. The timing resolution of each hodoscope is 120\,ps, resulting in a $\beta^{-1}$ resolution of 2.5\%. The threshold-type Cherenkov counter, using a silica aerogel radiator with refractive index $n = 1.03$, can reject electron and muon backgrounds by a factor of 12,000. The threshold rigidities for antiproton and antideuteron are 3.8\,GV and 7.6\,GV, respectively. In addition, a thin scintillator middle-TOF with timing resolution of 320\,ps is installed between the central tracker and the solenoid, in order to detect low-energy particles that cannot penetrate the magnet wall. 

BESS-Polar II provided an antiproton spectrum in the energy range from approximately 200\,MeV/$n$ to 3\,GeV/$n$~\cite{besspbar}; 
below 500\,MeV/$n$, this is the highest-precision antiproton measurement currently available. 
It also set an exclusion limit for antihelium of $1.0\cdot10^{-7} (\text{m}^2\text{s sr GeV}/n)^{-1}$ in the range of 1.6--14\,GV~\cite{2012PhRvL.108m1301A}, {based on the specific assumption that antihelium and helium have the same spectral shape. It needs to be noted that this assumption is generally not correct if the production mechanisms for helium and antihelium are different.}
The last published antideuteron results relied on the previous BESS flights, which took place between 1997 and 2000, setting an exclusion limit at 95\% confidence level of $1.9\cdot10^{-4}(\text{m}^2\text{s sr GeV}/n)^{-1}$ in the range of 0.17--1.15 GeV/$n$~\cite{Fuke:2005it}. 

An extended BESS-Polar II antiproton and the antideuteron analyses are currently ongoing using the middle-TOF that lowers the energy range to about 100\,MeV/$n$.

\subsection{The AMS-02 Experiment\label{s-ams}}

\begin{figure}
  \begin{center}
    \includegraphics[width=0.5\linewidth]{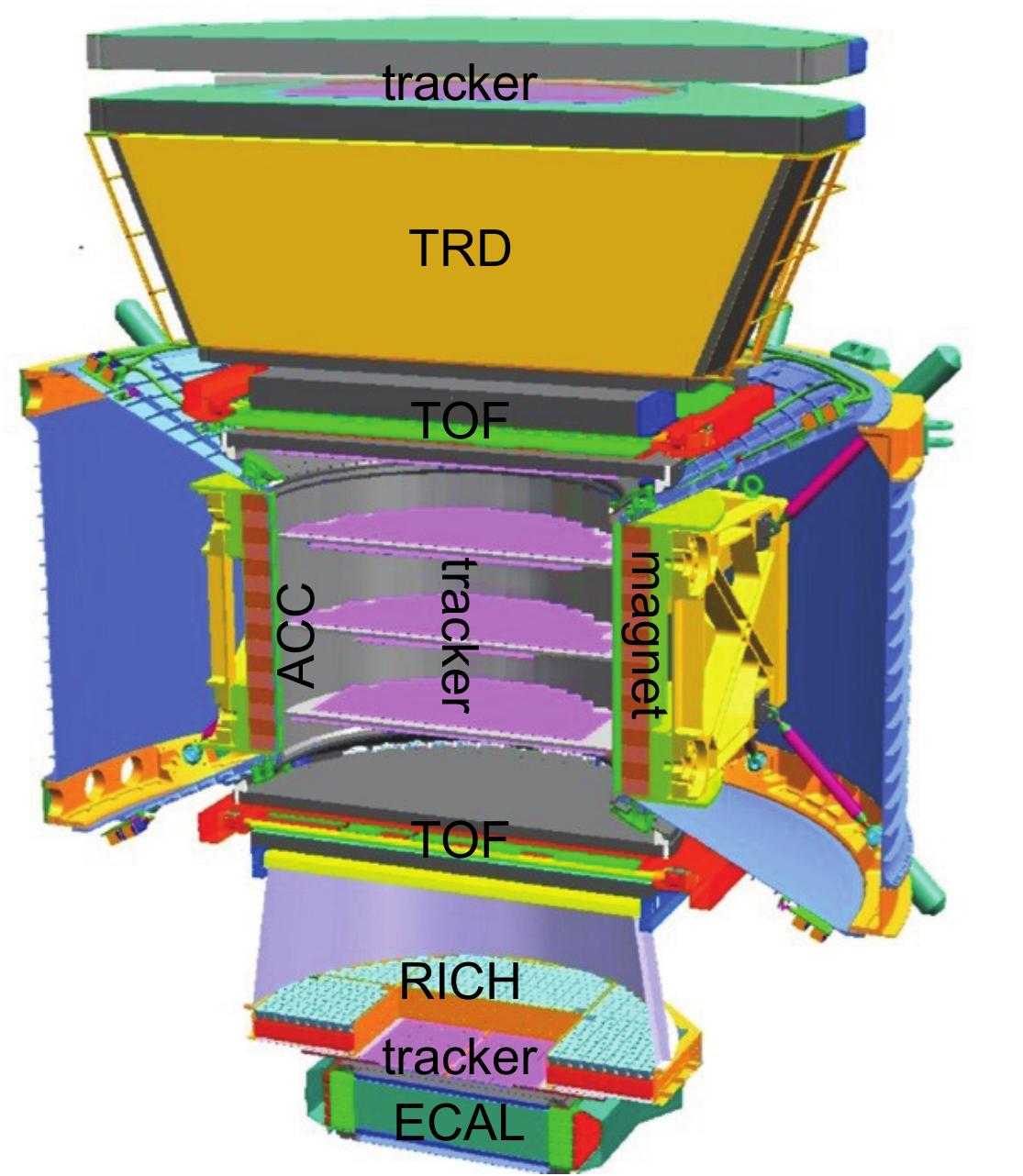}
    \end{center}
\caption{\label{f-fig2} AMS-02 is composed of a solenoid magnet and a series of detectors: the transition radiation detector, silicon tracker, anticoincidence counters, ring-imaging Cherenkov detector, electromagnetic calorimeter, and time-of-flight system. Reprinted with permission from~\cite{trackerupgrade}.}
\end{figure}

AMS-02 is a multi-purpose cosmic-ray detector that has been operating on the International Space Station (ISS) since May 2011. Thus far, it has recorded more than 140\,billion triggered events~\cite{ams4}, including more than 10\,billion protons and 100\,million deuterons. AMS-02 is planned to operate until the end of the lifetime of the ISS (at least 2024). In contrast to the high-statistics spectral measurements of other cosmic-ray species, the antideuteron and antihelium studies of AMS-02 are focused on a first-time discovery. 

AMS-02 follows the principle of typical magnetic spectrometer particle physics detectors, with particle identification that relies on combining signals from an array of sub-detectors, as shown in Fig.~\ref{f-fig2}. 
The transition radiation detector (TRD) is used to suppress low-mass particles such as electrons, pions, and kaons. The time-of-flight (TOF) system provides the main trigger and determines the velocity of the particle up to $\beta\approx0.8$. The particle momentum can be extracted from its trajectory in the approximately 0.15\,T solenoidal magnetic field. In the low-momentum region, multiple scattering becomes an important effect that limits the mass resolution. In the high-velocity region, two different types of ring imaging Cherenkov (RICH) counters are used (NaF and aerogel) for the velocity measurement. 

The AMS-02 collaboration has published the most precise antiproton spectrum in the range 1--450\,GV, based on $3.49\cdot10^5$ antiproton events and $2.42\cdot10^9$ proton events~\cite{PhysRevLett.117.091103}. The systematic uncertainty for charge confusion is $<1$\% below 30\,GV and 12\% at 450\,GV, and the uncertainty for the proton-carbon inelastic cross section is 4\% at 1\,GV and about 1\% above 50\,GV. By 2024,  more than 1,000,000 antiprotons will be collected in the range up to 525\,GV. As discussed further in Sec.~\ref{s-hebar}, AMS-02 has announced several candidate events with mass and charge consistent with antihelium~\cite{antihe,antihe2}. The analyses of heavy antinuclei, including antideuterons, are ongoing while more data are collected. At an estimated signal to background ratio of one in one billion, a detailed understanding of the instrument is required for these rare antinuclei searches. An equivalent of 100 billion of proton, deuteron and antiproton events in the rigidity range 0.5--100\,GV must be simulated. 

\subsection{The GAPS Experiment\label{s-gaps}}

\begin{figure}
  \begin{center}
    \includegraphics[width=0.8\linewidth]{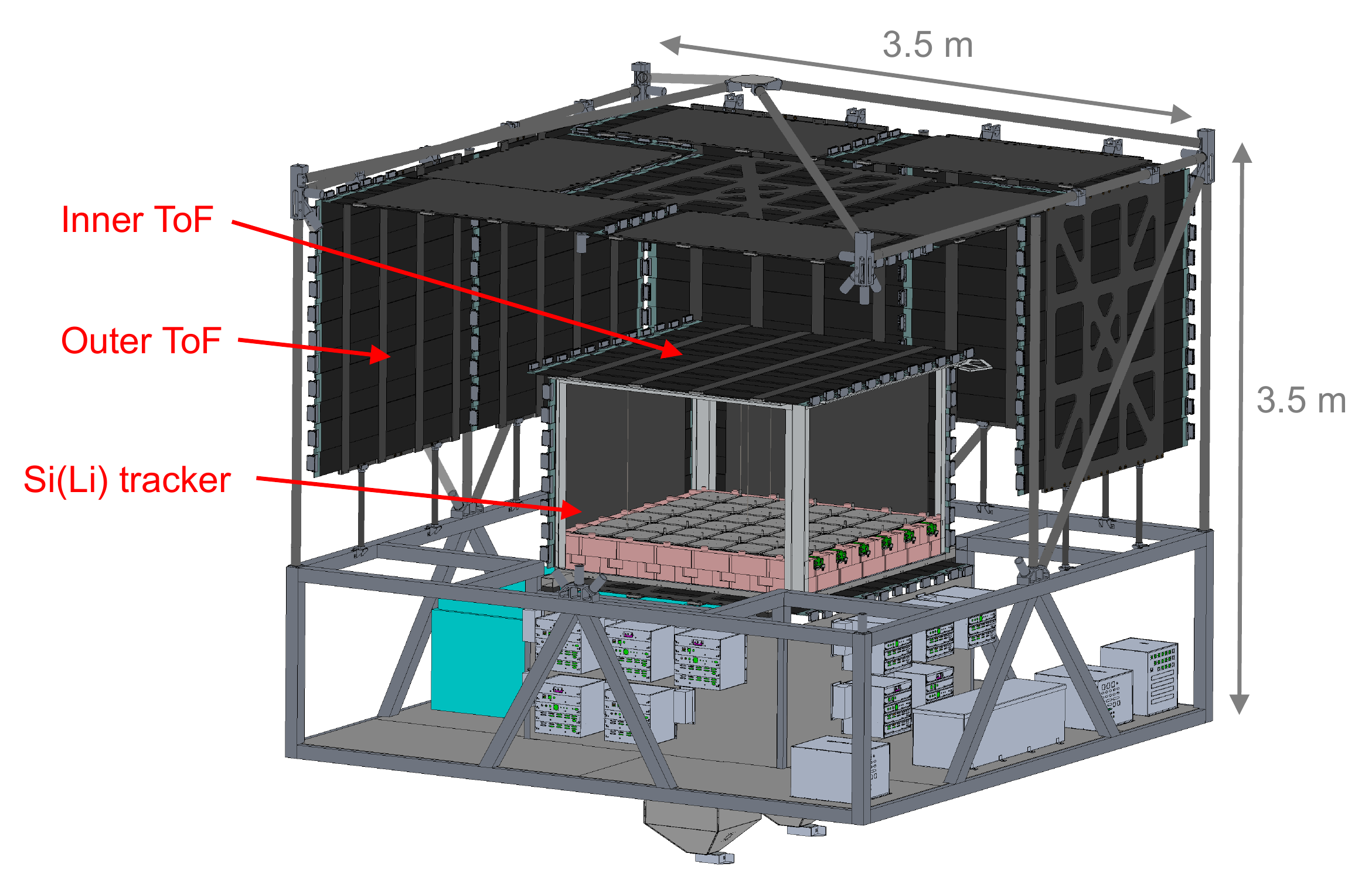}
    \end{center}
\caption{\label{f-fig3} The GAPS detection scheme relies on a tracker, which is composed of ten layers of Si(Li) strip detectors (only two layers shown), surrounded by a plastic scintillator time-of-flight system.}
\end{figure}

The General Antiparticle Spectrometer (GAPS) experiment is optimized specifically for low-energy ($<0.25$~GeV/n) cosmic-ray antinuclei~\cite{Aramaki2015}. The experiment, shown in Fig.~\ref{f-fig3}, consists of ten planes of semiconducting Si(Li) strip detectors surrounded on all sides by a plastic scintillator TOF. GAPS is scheduled for its first Antarctic long-duration balloon (LDB) flight in late 2021, with the baseline sensitivity to antideuterons projected after a total of three Antarctic LDB flights.

GAPS relies on a novel particle identification technique based on exotic atom formation and decay~\cite{Aramaki2015}. First, a low-energy antiparticle that has been slowed by the atmosphere passes through the TOF, which provides a high-speed trigger and measures particle velocity and $\text dE/\text dx$. It is further slowed by $\text dE/\text dx$ losses in the Si(Li) detector tracking system, eventually stopping inside the instrument. It then, with near unity probability, replaces a silicon shell electron to form an exotic atom in an excited state. This exotic atom then de-excites through auto-ionization and radiative transitions, emitting X-rays. These X-ray energies are uniquely determined by the antiparticle and silicon reduced mass and atomic numbers. Finally, the antiparticle annihilates with the silicon nucleus, producing a nuclear star of pions and protons. The simultaneous occurrence in a narrow time window of X-rays of characteristic energy and nuclear annihilation products with measured multiplicity provides an enormously constraining signature to distinguish antiparticles and to suppress non-antiparticle background. The main challenge for identifying antideuterons is the rejection of the dominant antiproton background, which is reached by combining different event signatures, including the total energy depositions, stopping depth, number of secondary tracks, and X-ray energies.

This exotic atom detector design yields a large grasp compared to typical magnetic spectrometers, and allows for the identification of antiproton, antideuteron, and antihelium cosmic rays. The TOF scintillator paddles are coupled to SiPMs on both ends and reach an end-to-end timing resolution of about 300\,ps, resulting in a velocity resolution of less than 10\% and longitudinal position resolution of less than 10\,cm. The TOF also  generates the trigger for the readout, which requires a hierarchical trigger system to focus on antiparticle triggers with a rate of less than 500\,Hz. The tracking system consists of 10\,layers with 10\,cm separation. Each layer is composed of $12\times12$ Si(Li) detectors of 10\,cm diameter and 2.5\,mm thickness, segmented into eight strips. The detectors have a demonstrated FWHM resolution of $<4$\,keV for 60\,keV X-rays at an operating temperature of $-40^\circ$C~\cite{2018NIMPA.905...12P,2019arXiv190600054R,2019NIMPA.94762695K}. In-flight, the detectors will be cooled with a passive oscillating heat pipe approach, which was tested on two prototype test flights \cite{Okazaki2014b,Okazaki2014a}.

GAPS will provide a precision antiproton spectrum for the first time in the low-energy range below $0.25$~GeV/$n$. As discussed in Sec.~\ref{s-dbar}, GAPS will provide a sensitivity to antideuterons that is almost two orders of magnitude better than the current BESS limits. Though the instrument is optimized for antideuteron sensitivity, the exotic atom detection technique is also sensitive to antihelium signatures. Due to the higher charge, the antihelium analysis suffers less from antiproton backgrounds, which allows for a competitive antihelium sensitivity in the low-velocity range ($\beta<0.6$).

\section{Impact of Current Cosmic Antinuclei Measurements\label{s-impact}}

In recent years, several anomalies in cosmic-ray spectra, for example in positrons~\cite{ams02first,ams4}, $\upgamma$-rays~\cite{Hooper:2011ti,TheFermi-LAT:2015kwa,Daylan2016,TheFermi-LAT:2017vmf,Hooper:2018fih,2019ApJ...880...95K}, and antiprotons~\cite{PhysRevLett.118.191101,PhysRevLett.118.191102,Reinert:2017aga,Cui:2018klo,Cuoco:2019kuu,Cholis:2019ejx}, have offered the tantalizing possibility of evidence of dark matter (DM) interactions. 
However, any possible DM interpretation must be consistent with all observed cosmic-ray spectra and remain {statistically} significant given the systematic uncertainties, in particular from cosmic-ray production and propagation. 
This section reviews the DM interpretation of recent antiproton measurements, the reports of cosmic antihelium events by AMS-02, and the implications for ongoing antideuteron searches. 

\subsection{High-statistics Antiproton Spectra\label{s-pbar}}

\color{black}
The most recent measurement of the cosmic-ray antiproton flux by AMS-02 contains about $3.5\cdot10^5$ events over a rigidity range of 0.4 to 400\,GV~\cite{PhysRevLett.117.091103}. 
Such a high-statistics spectrum has significantly increased the sensitivity to any DM signal, providing leading constraints on thermal WIMP DM across a wide mass range and uncovering a possible excess of approximately 10 to 20\,GeV antiprotons that is consistent with canonical WIMP annihilation signatures~\cite{PhysRevLett.118.191101, PhysRevLett.118.191102, Cui:2018klo,Cholis:2019ejx, Cuoco:2019kuu}. 
Given that this excess is at the level of about $10\%$ of the total flux, any interpretation relies on precise understanding of the systematic uncertainties. 
However, the persistence of such a possible DM signature certainly further motivates searches for other cosmic antinuclei species.

\subsubsection{Leading Dark Matter Constraints from Antiprotons}

\begin{figure}
\centering
\includegraphics[height=0.22\textheight]{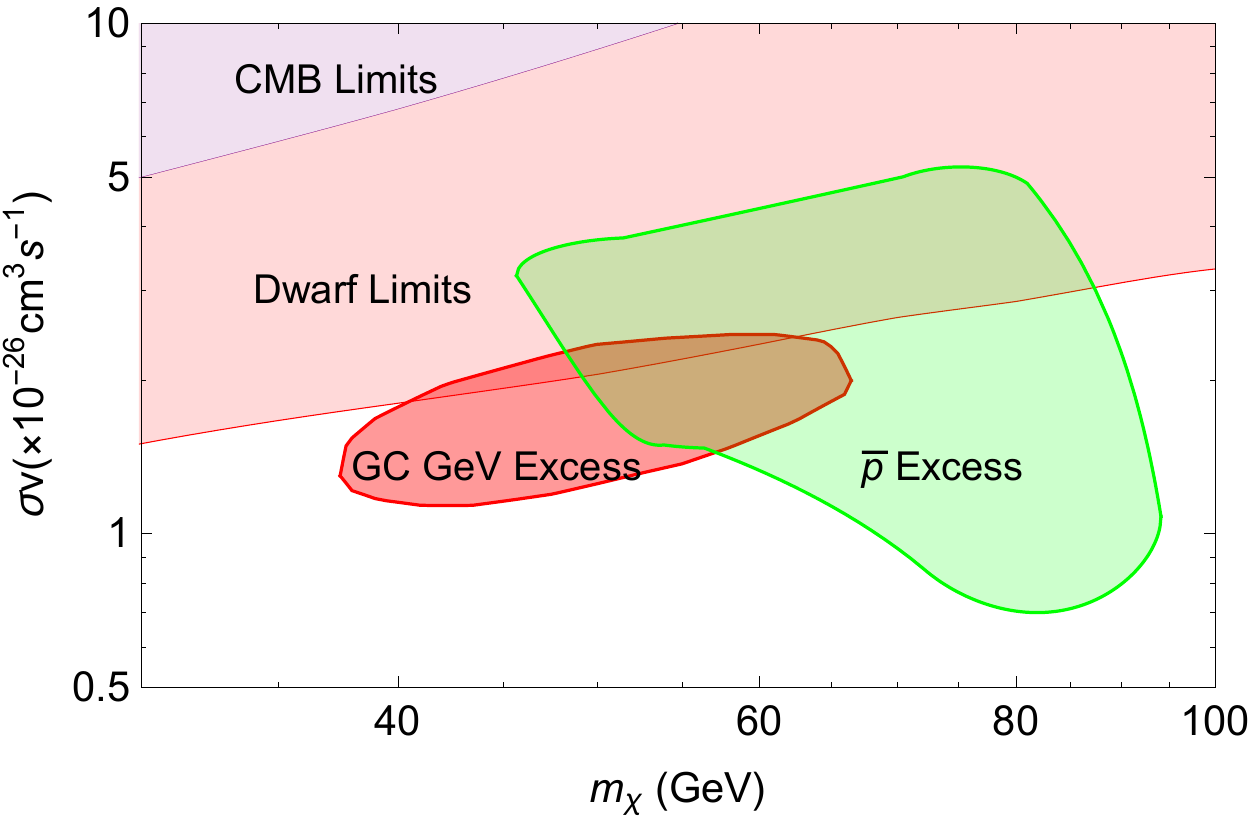}
\caption{\label{f-fig4} The regions of DM parameter space favored (within 2$\sigma$) by the AMS-02 antiproton spectrum (green closed) and the Galactic Center gamma-ray excess (red closed) for the case of annihilations into $b\bar{b}$, in the plane of velocity averaged annihilation cross section $\langle \sigma v \rangle$ and DM mass $m_\mathrm{DM}$. Reprinted with permission from ~\cite{Cholis:2019ejx}.}
\end{figure}

Across most of the energy range, the shape and normalization of the antiproton spectrum is in good agreement with the expectations of standard cosmic-ray production and transport models, allowing for stringent constraints on a wide range of DM models up to masses of about 1\,TeV.
Interpreted in terms of upper limits on DM annihilating hadronically into purely $b\bar{b}$ excludes a thermal annihilation cross section for DM masses below about 50\,GeV and in the range of 150 to 500\,GeV, even for conservative propagation scenarios~\cite{PhysRevLett.118.191102,Cholis:2019ejx}. 
{As shown in Fig.~\ref{f-fig4}, the favored region for the explanation of the cosmic-ray antiproton excess in terms of DM annihilation into $b\bar{b}$ in $\langle\sigma v\rangle$-$m_\chi$ space extends down by about a factor of 4 at 80\,GeV in comparison to the gamma-ray limits from dwarf spheroidal galaxies~\cite{Ackermann2016}.}
For large DM masses, the antiproton spectrum gives leading constraints across a variety of possible annihilation channels~\cite{Cuoco2017b}.  
Above 200\,GeV,  limits for DM annihilation into quarks, gluons, neutrinos, and gauge and Higgs bosons are stronger than the gamma-ray dwarf spheroidal limits, and for DM annihilation into charged leptons, flavor-independent limits are competitive to those from dwarfs for DM masses above about 2\,TeV. 
These limits are robust to conservative assumptions of the antiproton production cross sections and solar and Galactic propagation, and unlike gamma-ray limits from dwarf spheroidal and Galactic Center observations, are relatively insensitive to the choice of DM density profiles. 

\subsubsection{A Possible Dark Matter Signal in Antiprotons?}

\begin{figure}
\centering
\includegraphics[height=0.3\textheight]{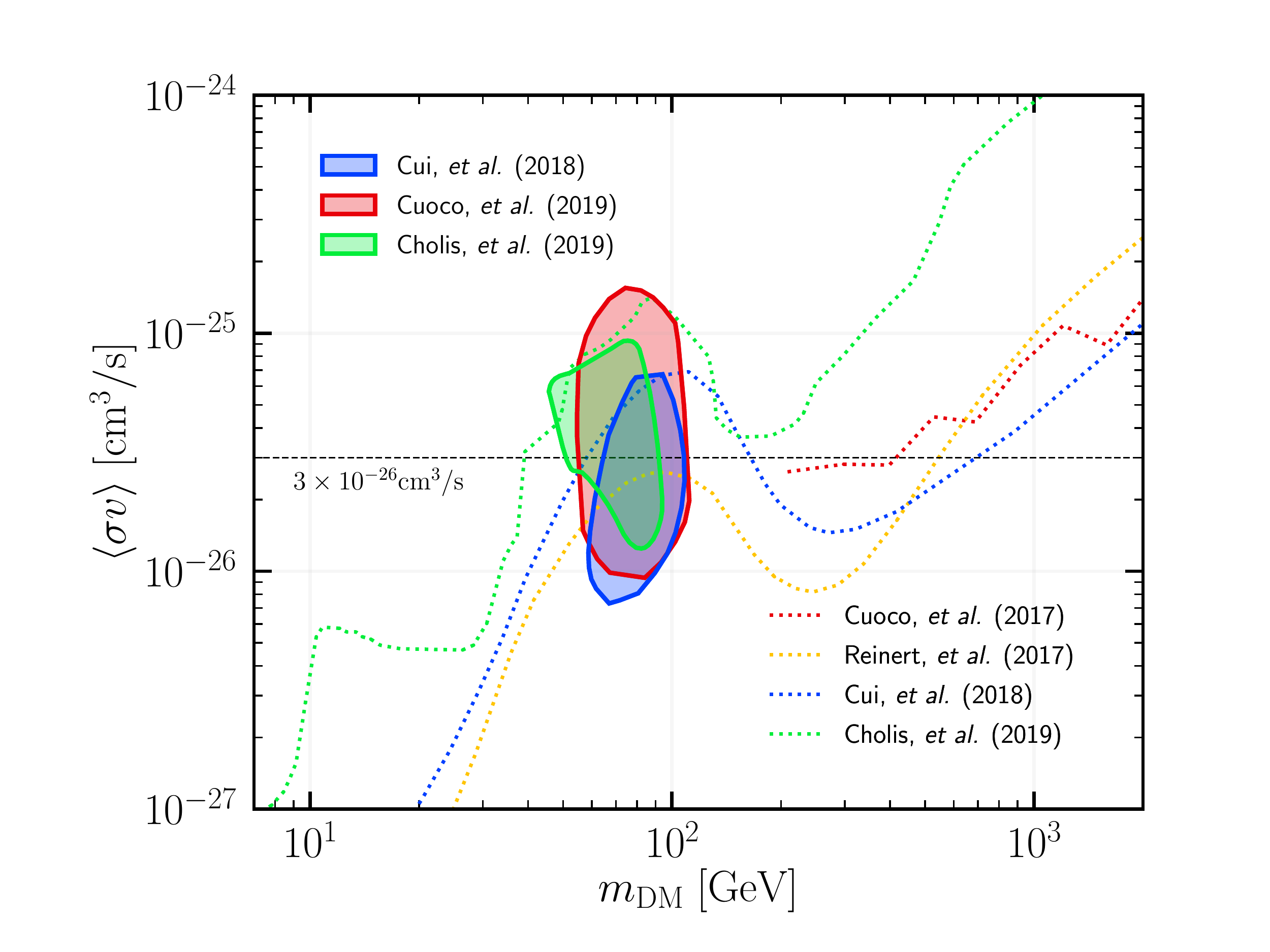}
\caption{\label{f-fig5} The regions of DM parameter space favored by recent analyses of AMS-02 antiproton data for the case of annihilations into $b\bar{b}$, in the plane of velocity averaged annihilation cross section $\langle \sigma v \rangle$ and DM mass $m_\mathrm{DM}$. The contours of Cuoco, et al. 2019~\cite{Cuoco:2019kuu} (red) and Cholis, et al. 2019~\cite{Cholis:2019ejx} (green) refer to the 2$\sigma$ best-fit region in the frequentist interpretation, while the contour of Cui, et al. 2018~\cite{Cui:2018klo} (blue) corresponds to a 95\% C.L. in Bayesian interpretation.  
The exclusion limits from four recent analysis are also shown~\cite{Cuoco2017b,Reinert:2017aga,Cui:2018klo,Cholis:2019ejx}. 
We note that the analysis of Reinert, et al. 2017~\cite{Reinert:2017aga} does not find a significant DM signal, but the derived limit weakens at  $m_\mathrm{DM} \approx80$~GeV such that all indicated DM signal regions lie partly below this limit. 
For a thermal WIMP one would expect a annihilation cross section at the order of $\langle \sigma v \rangle \approx3 \times 10^{-26}\,\mathrm{cm^3/s}$ (black dashed line).
The signal regions and limits have been rescaled to a local DM density of $\rho_0 = 0.3\,\mathrm{GeV/cm^3}$. Note that the different analyses use slightly different values for the halo half-height which affects the value of $\langle \sigma v \rangle$.}
\end{figure}

At low energies, an excess of 10--20\,GeV cosmic-ray antiprotons has been identified~\cite{PhysRevLett.118.191101, PhysRevLett.118.191102, Cui:2018klo,Cholis:2019ejx, Cuoco:2019kuu}, as shown in Fig.~\ref{f-fig5}. 
However, the significance of this possible DM signal of course depends on the interpretation of the theoretical and experimental systematic uncertainties, in particular careful treatment of any energy correlations. 

Interpreted as a signal of hadronically annihilating DM, this corresponds to a DM mass of approximately 40-130\,GeV and a thermal annihilation cross section of about $3\cdot10^{26}$\,cm$^3$/s.
Intriguingly, this signal is consistent with a range of DM models that account for GC excess signal~\cite{Hooper:2010mq, Hooper:2011ti, Abazajian:2012pn, Gordon:2013vta, Daylan2016, Calore:2014xka, TheFermi-LAT:2015kwa} (see, e.g. Fig.~\ref{f-fig4}) and many beyond the Standard Model scenarios~\cite{2017arXiv171106460C, Clark:2017fum, Cirelli:2016rnw, Cline:2017lvv, Jia:2017kjw, Arcadi:2017vis, Li:2017nac, Escudero:2017yia}. 

In addition, as the interpretation of the excess in terms of hadronic annihilation implies substantial coupling between DM and quarks, it is natural to test compatibility with constraints from direct detection. 
Using the framework of effective operators for fermionic DM, which is only valid if the mediator is much heavier than the DM, generic operators with an unsuppressed annihilation cross section (s-wave) but with a suppressed (either by velocity or momentum) DM-nuclei scattering interaction can reproduce both the antiproton signal and the combined bounds from PandaX-II, LUX, and XENON1T~\cite{Cui:2018nlm}. 

A dominant uncertainty on the overall spectrum comes from the modeling of the {interstellar} and {heliospheric} propagation.  
Using the observed spectra of protons, helium, and the antiproton-to-proton ratio from AMS-02 and the extra-solar proton and helium spectra from VOYAGER, Cuoco et al. ~\cite{Cuoco:2019kuu} conclude that the antiproton-to-proton ratio favors an excess at the 3.1$\sigma$ level. 
In addition, Cholis et al.~\cite{Cholis:2019ejx} use three representative models for cosmic-ray injection and transport, each of which provides a good overall fit to the observed cosmic-ray proton, helium, carbon, and boron-to-carbon ratio spectra up to 200\,GV, and conclude that the excess remains at a  $4.9\sigma$ significance after accounting for solar modulation, Galactic propagation, and antiproton production cross section. 
This significance reduces slightly, to $3.3\sigma$, while improving the fit to the antiproton-to-proton ratio at energies above 100\,GeV, if allowing for acceleration of secondary antiprotons in supernova remnants. 

{The cross section of antiproton production in interactions of cosmic-ray protons and helium with interstellar gas}
is also a significant source of overall uncertainty. 
At low energies the uncertainty from the antiproton production cross sections increases, conservatively reaching a level of 10\% to 20\%~\cite{Korsmeier:2018gcy}. 
The possible DM signal remains significant when accounting for this uncertainty by either propagating the error to the {calculated flux} using a covariance matrix or by performing a simultaneous fit of the antiproton flux and cross section parameters~\cite{Cuoco:2019kuu}, or by allowing the cross section to vary within its $3\sigma$ uncertainties while exploring the antiproton spectral fit~\cite{Cholis:2019ejx}. 
This is in contrast to a previous analysis~\cite{Reinert:2017aga}, which found the cross section uncertainty reduces the signal significance to about $1\,\sigma$.
{It is important to keep in mind that there are several small differences in the setup of all four analyses, which might be the origin of the discrepancy in significance.}

As the spectral shape {of antiprotons from DM annihilations} is very different from the astrophysical secondary or tertiary spectra, the significance of the potential DM signal may critically depend on correlations in the uncertainties of both the measurement and the predicted spectrum.
Short-range correlations, in particular, can affect the significance of {narrow spectral} features like the ones expected from DM. 
So far, AMS-02 has only provided absolute, not correlated, systematic uncertainties for their measurements. 
The reduced chi-squared for fits to the proton and helium spectra are much less than one, which could indicate that these systematic uncertainties are either over-estimated or significantly correlated. 
In the absence of official estimates of the correlations, a data-driven likelihood approach has been used to constrain long-range, short-range, and uncorrelated uncertainties in the measurement, with the DM signal still being significant at the level of about $3-5\,\sigma$~\cite{Cuoco:2019kuu}. 
This is in contrast to the work of Boudaud et al.~\cite{Boudaud:2019efq}, who define full energy-dependent correlations of the uncertainties due to  a benchmark transport model defined by AMS-02 B/C data~\cite{Genolini:2019ewc}, the antiproton production cross sections, and measurement effects quoted by the AMS-02 collaboration. 
This work finds that accounting for the statistical errors on the measurement and the energy-dependent correlations of either the theoretical predictions or the measurement alone can eliminate the significance of the DM signal.

\subsubsection{Prospects for Improved Antiproton Measurements}

GAPS will measure a precision antiproton spectrum in a low-energy region currently inaccessible to any experiment.
With one flight, GAPS is expected to identify $>1000$ antiprotons in the energy range $E < 0.25$\,GeV/n.
Such a high-statistics spectral measurement has already been demonstrated to yield leading DM constraints at higher masses. 
Since the antiproton spectrum from DM annihilations shifts towards lower kinetic energies with decreasing DM mass, the precision measurement {in} the GAPS energy range offers new phase space for probing light DM models and primordial black holes~\cite{Page:1976wx,Maki:1995pa,2002A&A...388..676B}.
The GAPS ultra-low energy antiproton measurement will be sensitive to light neutralinos, gravitinos, and Kaluza-Klein DM for a conservative range of propagation parameters~\cite{Aramaki:2014oda}. 
The GAPS antiproton measurement will also allow for sensitive studies of systematic effects, in particular propagation of antinuclei in the {interstellar} and {heliospheric} environments. 

\subsection{Interpretation of Antihelium Candidates\label{s-hebar}}

Since its launch in 2011, AMS-02 has accumulated several billion events, whose composition is mostly dominated by protons and helium nuclei.
To date, AMS-02 has reported observation of eight antihelium candidate events with rigidity below 50\,GV, from a sample of 700 million selected helium events. 
Six of these candidates have mass in the range of antihelium-3, and two candidates have mass in the range of antihelium-4. 

Although these results have not yet been published in a peer-reviewed journal, 
should they be confirmed, the detection of cosmic antihelium would be a breakthrough discovery with immediate and considerable implications for our current understanding of cosmology. 
As the discovery of even a single antihelium-4 nucleus is challenging to explain in terms of known physics, confirmation with an independent detection technique, such as provided by GAPS, is essential. 
If taken seriously, these candidate antihelium events have broad implications for measurements of other cosmic-ray antinuclei species. 

Cosmic-ray antihelium has been identified as a potentially promising indirect detection signature due to its extremely low astrophysical backgrounds, especially towards low energies. 
When comparing the expected astrophysical flux of antiprotons,  antideuterons, antihelium-3, and antihelium-4,  each subsequent nucleus suppresses the flux by a factor of $10^4$ to $10^3$ due to coalescence {probability} effects.  
Of course, searching for a DM signature in antihelium is challenging, as the {same} coalescence effects also suppress any possible DM signal. 
In addition, for a given DM model, the predicted cosmic-ray antihelium flux is expected to be strongly correlated with that of cosmic-ray antiprotons, and so must satisfy the precise bounds discussed in Sec.~\ref{s-pbar}.

Previous work on the potential of antihelium searches concluded that 
realistic fluxes of antihelium-3 from DM are outside the sensitivity reach of GAPS and AMS-02, even assuming a larger coalescence momentum for antihelium than antideuterons~\cite{Carlson:2014ssa}, 
and that for realistic predictions, the experimental sensitivity must be improved by a factor of  approximately 500 to 1000 for even a single antihelium-3 detection~\cite{Cirelli:2014qia}.

\begin{figure}
\centering
\includegraphics[width=0.5\linewidth]{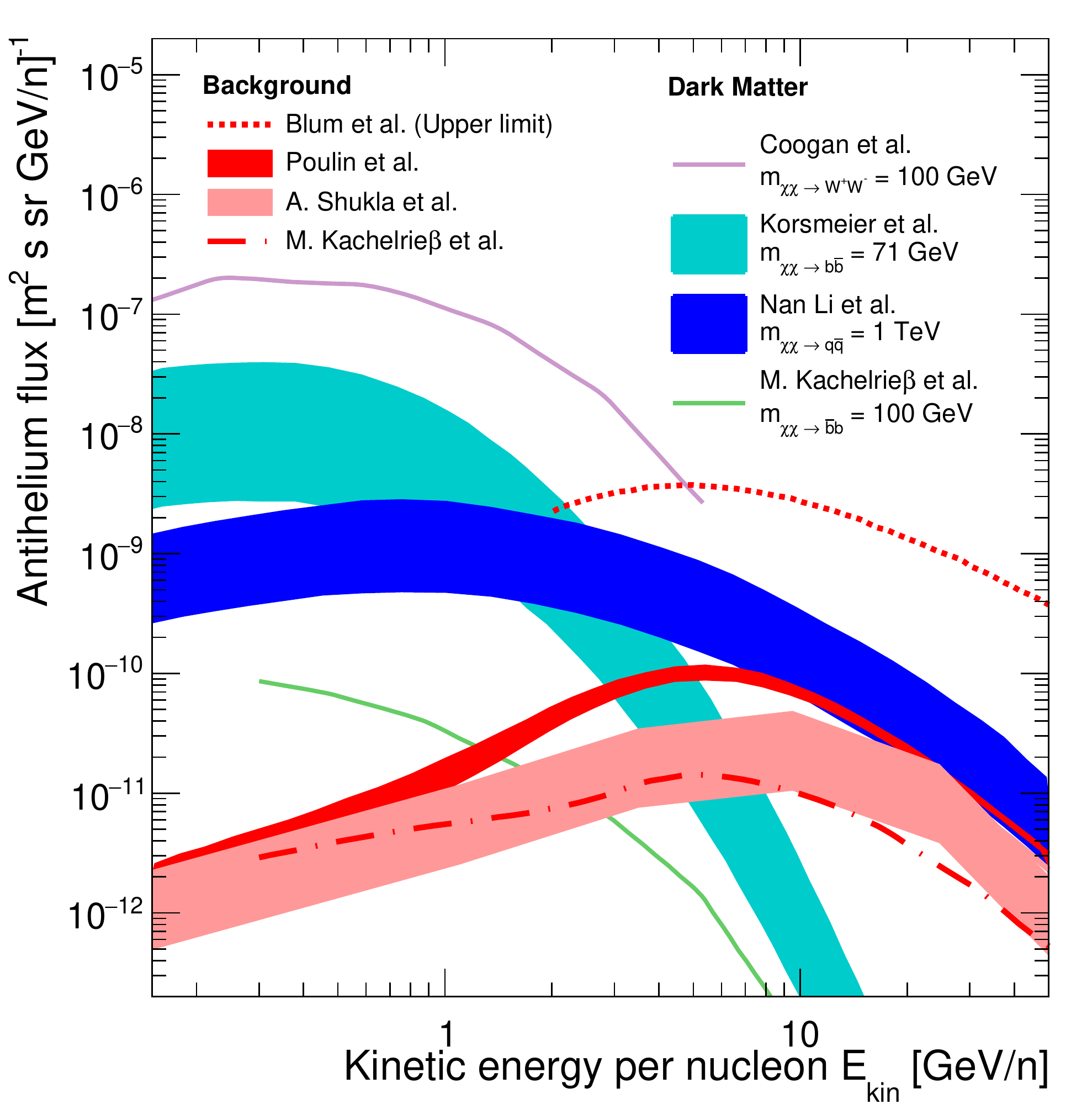}
\caption{\label{f-fig6} Antihelium-3 flux predictions from different DM model and astrophysical background calculations~\cite{Korsmeier:2017xzj,Coogan:2017pwt,Blum:2017qnn,Li:2018dxj,2018arXiv180808961P, Kachelriess:2020uoh, uhantihe3}. The error bands illustrate uncertainties in the coalescence momentum, but also include propagation uncertainties.}
\end{figure}

In light of the recent announcements of antihelium candidate events, the predictions for astrophysical and DM production of antihelium-3 have been revisited. 
In particular, optimistic coalescence scenarios have been investigated that can boost the expected antihelium signal enough to be within the detection range of AMS-02 without conflicting with antiproton observations. 
Crucially, the interpretation of these events as either DM or astrophysical depends on their energies, which have not been reported for all candidate events. 
The detection of antihelium-3 can be compatible with antiproton bounds if due to DM annihilating into purely $b\bar{b}$ and assuming a maximal coalescence momentum, but only if the events are distributed across the full sensitive energy band of AMS-02. {However, this requires DM annihilation cross sections} that are in tension with the gamma-ray dwarf spheroidal limits~\cite{Coogan:2017pwt}. 
For events in the higher energy region of the AMS-02 sensitivity, optimistic coalescence and hadronization scenarios have been developed~\cite{Herms:2016vop, Korsmeier:2017xzj,Blum:2017qnn, Li:2018dxj} that predict a secondary astrophysical flux of antihelium-3.
Fig.~\ref{f-fig6} shows a wide range of fluxes for both DM induced and astrophysical antihelium-3 flux.
However, these scenarios may conflict with the coalescence now measured by ALICE, which is close to constant over a wide range of high center-of-mass energies (0.9--7\,TeV), and with direct measurements in heavy ion collisions in the energy range 0.4--2.1\,GeV/n, which predict an even lower value of the coalescence momentum~\cite{2018arXiv180808961P}. 
Thus explanations of this signal purely in terms of DM seems to require specifically-tuned models, such as theories in which a new particle carries baryon number, favoring the production of nuclear bound states over the decay into individual nucleons and allowing the flux of antihelium from DM to dominate over the flux of antiprotons from DM~\cite{Heeck:2019ego}.

{In contrast to the case of antihelium-3, even optimistic coalescence scenarios cannot account for the detection of even a single cosmic antihelium-4 nucleus from either astrophysical or DM origins. 
Despite the large uncertainties in production and propagation models, it is impossible to account for the observed flux of antihelium-4 with secondary production, and generic DM scenarios face similar difficulties. 
Therefore, the discovery of a single antihelium-4 was long known as being a smoking gun of the presence of antiobjects in the galaxies. 
Large-scale antimatter over-densities face constraints from gamma-ray observations of the extragalactic background~\cite{vonBallmoos:2014zza}, but local over-densities are still possible. 
Hence, the observed rate of cosmic antihelium could indicate the existence of local antistars or anticlouds~\cite{2018arXiv180808961P}. 
Interestingly, these objects are not expected to produce antideuterons as they are burned in nuclear reactions within the stars. Antistars and other antiobjects can be produced from baryon isocurvature modes at small scales. These arise for instance in some modified version of the Affleck-Dine baryogenesis~\cite{Affleck:1984fy}, as suggested in Ref.~\cite{Dolgov:1992pu} and revisited recently in Ref.~\cite{Blinnikov:2014nea}. In these models, the dynamics of a complex scalar field lead to the observed homogeneous baryon asymmetry, but also allow inhomogeneous pockets with large baryon or antibaryon number. 
In particular, antistars whose main material is antihelium-4, converted into antihelium-3 via spallation in the dense surrounding environment, would be naturally free of matter at their heart. Thus annihilation would be limited to their surface, allowing survival until today and reducing the expected signature in other cosmic-ray species.
These models also allow to connect DM to the presence of antiobjects, as a simultaneous production of a DM component in the form of primordial black holes or other compact objects is expected. Detection or exclusion of such a DM component through other astrophysical probes could, therefore, help to understand or constrain the origin of the antihelium candidate events seen by AMS-02.}

\subsection{Implications for Antideuteron Detection\label{s-dbar}}

The prospects for cosmic antideuteron detection must now be considered in light of the precise measurements of the cosmic-ray antiproton spectrum and increased theoretical work on production cross sections and coalescence prompted by the candidate cosmic-ray antihelium events. 
As both GAPS and AMS-02 should soon provide sensitivity to the relevant flux ranges of antideuterons, investigation of the possible antiproton and antihelium signals via this new channel is both compelling and timely. 
Antideuterons can also provide sensitivity to DM signatures that would not be visible in antiprotons, due to the abundant antiproton secondary background, or in antihelium, due to suppression of heavier antinuclei formation {in the process of} coalescence. 
Perhaps more importantly, the history of cosmic-ray experiments indicates that {it is important to }
remain open to serendipitous, unexpected results when first probing {higher sensitivity grounds}.

Investigating the associated production of antideuterons, whose secondary production is strongly kinematically suppressed at low energies, is the most direct option to cross-check the possible excess of antiprotons. 
As mentioned in Sec.~\ref{s-pbar}, the generic DM models that can account for the antiproton signal are also compatible with the long-standing puzzle of excess gamma-rays from the Galactic Center.
Antideuterons are thus uniquely suited to probe both of these {potential} signals. 

The whole parameter space {probed} by the antiproton signal~\cite{Cuoco:2017c} would produce a detectable signal in GAPS for a feasible range of Monte Carlo and analytic coalescence models, coalescence momenta, and {interstellar and heliospheric} propagation parameters~\cite{Korsmeier:2017xzj}. 

\begin{figure}
\centering
\includegraphics[width=0.6\linewidth]{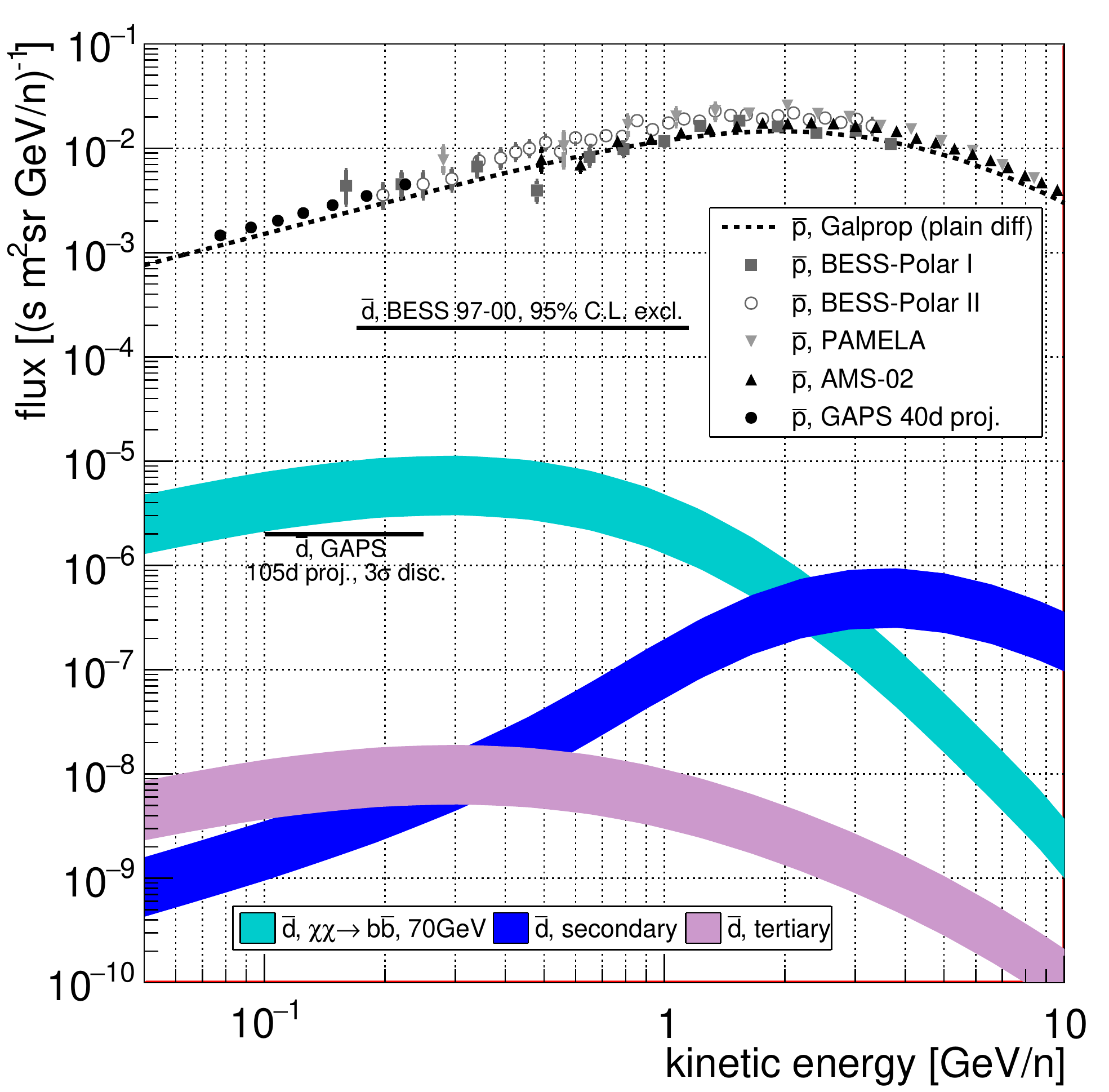}
\caption{\label{f-fig7}
Antiproton flux data from AMS-02~\cite{PhysRevLett.117.091103}, BESS-Polar I/II~\cite{Abe:2008sh,besspbar}, and PAMELA~\cite{2010PhRvL.105l1101A}, as well as projections for the GAPS~\cite{Aramaki:2014oda} antiproton flux measurements after 40~days, in comparison with the GALPROP plain diffusion prediction~\cite{Moskalenko:2001ya}.
Also shown are the predicted antideuteron flux corresponding to DM parameters indicated by AMS-02 antiproton signal, interpreted as annihilation into purely $b\bar{b}$~\cite{Cuoco:2017c,Korsmeier:2017xzj}), as well as the predicted secondary and tertiary astrophysical antideuteron flux. The anticipated sensitivity of GAPS~\cite{Aramaki2015} for a 3\,$\sigma$ discovery and the BESS 97--00 95\% C.L. exclusion limits are indicated~\cite{Fuke:2005it}.
Solar modulation is treated in the force-field approximation with a potential of 500\,MV. 
All antideuteron fluxes are derived in the analytic coalescence model with a coalescence momentum of 160\,GeV~\cite{Schael:2006fd} for the lower edge of the band and with a higher coalescence momentum of 248\,GeV~\cite{Acharya:2017fvb} for the upper edge of the band.}
\end{figure}

\begin{figure}
\centering
\includegraphics[width=0.6\linewidth]{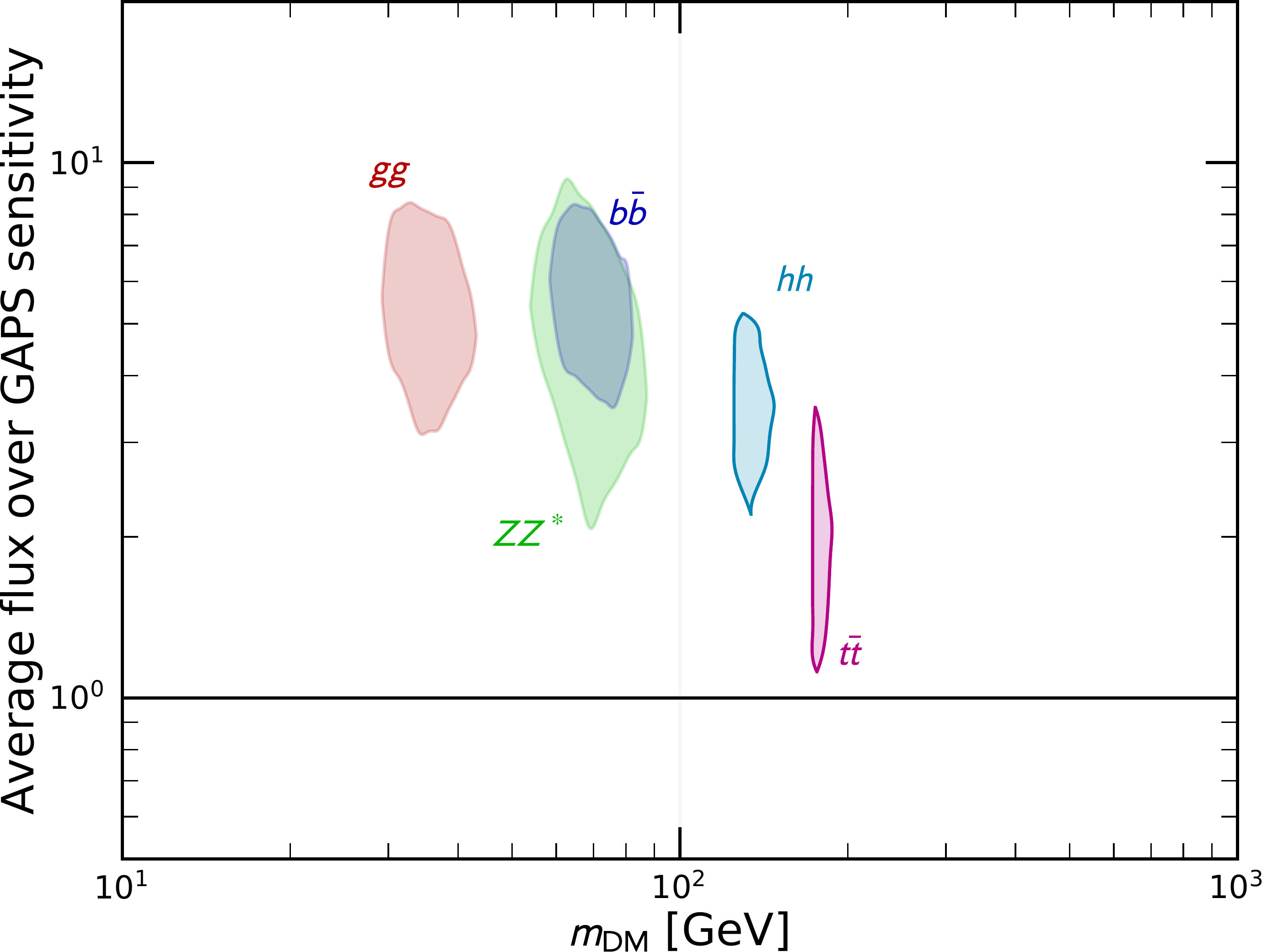}
\caption{\label{f-fig8}
The predicted antideuteron flux corresponding to the observed AMS-02 antiproton excess, interpreted as DM annihilation into purely $b\bar{b}$, two gluons, two $Z$-bosons, two Higgs-bosons, or $t\bar{t}$, divided by the expected GAPS sensitivity~\cite{Aramaki2015}.
A clear antideuteron signal is also expected in GAPS for all annihilation channels. 
For the $Z$-boson decay, one of the two bosons might be produced off-shell, which is denoted with a star superscript. 
The shaded regions correspond to the $2\sigma$ on the best-fit DM mass and annihilation cross section from antiproton data~\cite{Cuoco:2017c}.
Reprinted with permission from~\cite{Korsmeier:2017xzj}.
}
\end{figure}

The {normalization} of antideuteron flux depends most critically on the coalescence momentum. 
As shown in Fig.~\ref{f-fig7}, the peak antideuteron flux is clearly within the sensitivity range of GAPS, assuming the case of annihilation into $b\bar{b}$ and the coalescence momentum derived from measurements of antideuteron production via $Z$-boson decay~\cite{Schael:2006fd}. 
Since the initial state is not hadronic, antideuteron coalescence after DM annihilation can be considered closest to this case.  
Assuming a larger coalescence momentum of 248\,MeV, as recently suggested by the ALICE measurements~\cite{Acharya:2017fvb}, would increase the expected flux by a factor of four.
As shown in Fig.~\ref{f-fig8}, a clear antideuteron signal is also expected in GAPS for DM models that interpret the antiproton excess in terms of pure annihilation into two gluons, $Z$-bosons, Higgs-bosons, or top-quarks~\cite{Cuoco:2017c}. 
In comparison to the findings in ~\cite{Herms:2016vop}, which predicts lower antideuteron fluxes for DM models, it becomes apparent that understanding the interplay of the different uncertainties (Sec.~\ref{s-uncertain}) in antiproton flux as well as propagation models and coalescence is very important, highlighting again the importance of searching for heavier antinuclei to break the degeneracies.

The scenarios that could account for the antiproton excess can be considered more generally as illustrating a range of simple DM models that will be testable by antideuteron signatures, but invisible to collider and direct detection methods. 
Another generic case is that of DM in a hidden sector, which has extremely small mixing between light and dark sectors, such as dark photon DM~\cite{Randall:2019zol}. 
Such theories posit a fermionic DM particle with mass $m_\chi$ and a massive gauge boson mediator with mass $m_{A'} < m_\chi$. 
The DM is then permitted to annihilate into two mediators, which can in turn propagate and decay into Standard Model particles, such as $q\bar{q}$. 
The total rate of this process is insensitive to the small mixing between light and dark sectors, and thus if dark photon DM is a thermal relic, then only indirect detection methods can viably detect it. 
For a benchmark model with $m_\chi \approx 30-90$\,GeV and $m_{A'} \approx 10-50$\,GeV, the predicted flux of antideuterons would be within the sensitivity limit of GAPS~\cite{Randall:2019zol}.

Antideuterons also provide a crucial independent channel to constrain models of antihelium production. 
Any increase in coalescence probability, which is necessary to explain the antihelium-3 signal via astrophysical secondary or DM production, would necessarily also boost the antideuteron flux {in standard coalescence models.}
For example, both GAPS and AMS-02 are expected to detect a significant amount of antideuterons if the antihelium events originate from DM annihilating to, e.g. $b\bar{b}$~\cite{Coogan:2017pwt} ({though the reader is referred to the caveats} on any DM interpretation of the antihelium-3 signal discussed in Sec.~\ref{s-hebar}). 
In addition,  local antistars or anticlouds, such as could explain the observation of antihelium-4, would also produce a detectable flux of antideuterons in the low-energy band of GAPS~\cite{2018arXiv180808961P}. 

In summary, antideuterons remain a sensitive signature of a broad range of DM models.
These include a lightest supersymmetric particle (LSP) neutralino~\cite{Donato:2008yx}, a right-handed Kaluza-Klein neutrino of extra-dimensional grand unified theories (LZP)~\cite{Baer:2005tw}, a decaying LSP gravitino~\cite{Dal:2014nda}, a DM particle with gluonic decay channels~\cite{Cui:2010ud}, or candidates in the next-to-minimal supersymmetric model (NMSSM)~\cite{Cerdeno:2014cda}. Depending on details for antideuteron formation and cosmic-ray propagation models, the experimental sensitivities reach well below the thermal relic cross section for annihilation into light quarks for DM with mass below 20--100~GeV~\cite{Fornengo:2013osa}. A detectable antideuteron signal may also result from annihilation in certain heavy (5--20\,TeV) WIMP models, if one assumes a cosmic-ray propagation scenario with a larger halo size~\cite{Brauninger:2009pe}. 
In addition, models that combine TeV-mass, pure-Wino DM with Sommerfeld enhancement mechanisms can result in accessible antideuteron signals~\cite{Hryczuk:2014hpa}. 

\section{Uncertainties for Cosmic Antinuclei Flux Predictions\label{s-uncertain}}

The interpretation of cosmic-ray antinuclei data is impacted by the limited knowledge of antinuclei production in interactions of primary cosmic rays with the interstellar {gas} and of cosmic-ray propagation parameters. This section discusses these systematic uncertainties and gives an outlook on how they can be reduced.

\subsection{Antiproton Production Cross Section\label{s-antipxs}}

\begin{figure}
\centering
\includegraphics[height=0.445\linewidth]{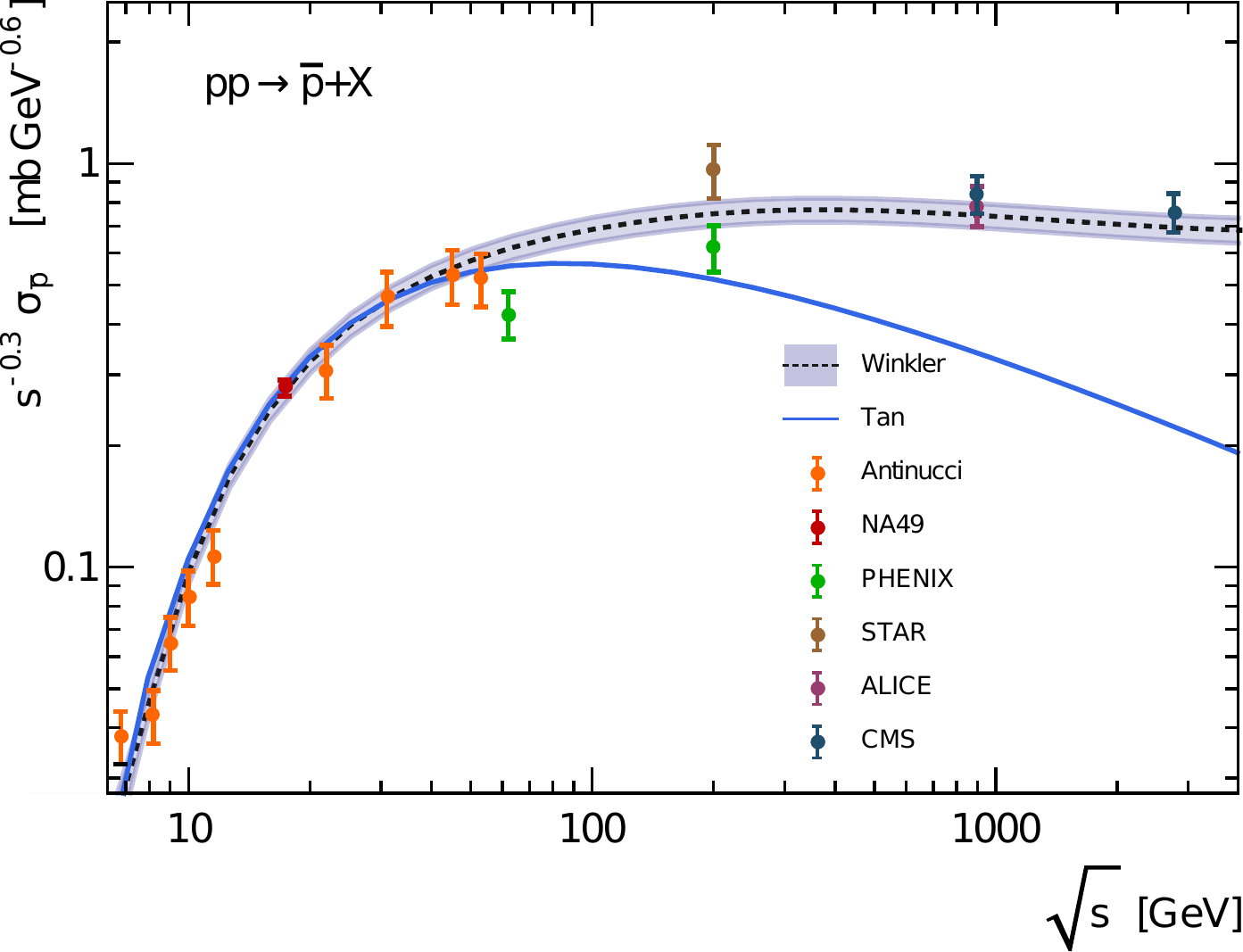}
\caption{\label{f-fig9}Antiproton production cross section in proton-proton interactions~\cite{Winkler:2017xor}.}
\end{figure}

{ The cross section for antiprotons to be produced by cosmic-ray interactions with the interstellar medium is key to interpreting cosmic antinuclei measurements. For example, }
AMS-02 data revealed that the cosmic-ray antiproton flux { at energies above approximately 100\,GeV} is harder than was anticipated. Although this prompted a number of different interpretations, including antiprotons produced by heavy DM~\cite{2015arXiv150405937H,Chen:2015cqa}, it was found that the antiproton production cross section in $p$--$p$ collisions at higher energies was underestimated. A new parameterization~\cite{Winkler:2017xor}, especially at energies above $\sqrt{s}>100$\,GeV, considering PHENIX~\cite{Adare:2011vy}, STAR~\cite{Abelev:2006cs,Abelev:2008ab}, CMS~\cite{Khachatryan:2011tm,Chatrchyan:2012qb,Zsigmond:2012vc,H.VanHaevermaetfortheCMS:2016uir}, and ALICE~\cite{Aamodt:2011zza,Aamodt:2011zj} data (Fig.~\ref{f-fig9}), together with a new parameterization of the boron-to-carbon ratio (see Sec.~\ref{s-prop}) resolved the discrepancy in the predicted and measured antiproton flux. The uncertainties on antiproton production from protons interacting with heavier nuclei are even larger than those from $p$--$p$ interactions, { because very few direct measurements exist and these cross sections are instead} calculated by rescaling the $p$--$p$ cross sections. 

At lower energies, new $p$--$p$ data ($\sqrt{s}=7.7, 8.8, 12.3, 17.3$\,GeV) became available from NA61/SHINE in 2017~\cite{Aduszkiewicz:2017sei}. In addition, the first antiproton production cross section in $p$--He collision from LHCb at $\sqrt{s}=110$\,GeV was published\cite{Aaij:2018svt}. Still, cross section uncertainties in the energy range of AMS-02 are at the level of 10--20\%, with higher uncertainties for lower energies. For energies lower than the AMS-02 range, relevant for the GAPS experiment, a significant uncertainty on the source term from cross section normalization and shape exist. A recent study highlighted that, in particular, future measurements at low center-of-mass energies ($<7$\,GeV) could improve these antiproton flux uncertainties~\cite{Donato:2017ywo}. Furthermore, it was found that when trying to fit the cosmic-ray antiproton spectrum and allowing the cross section and the cosmic-ray propagation parameters to vary the significance of the DM interpretation {of the excess in the flux at 10--20\,GeV} was only slightly affected by the uncertainty of the antiproton production cross section~\cite{Cholis:2019ejx}. 
{Nevertheless, improving on antiproton cross section measurements still remains very relevant for a precision understanding and the antinuclei formation discussed in the next section.}

\subsection{Antinuclei Formation\label{s-antinuclprod}}

For heavier antinuclei made of multiple antinucleons, it is important to note that typically every production process should also produce antiprotons in much higher quantity, with each additional antinucleon reducing the yield by a factor of approximately 1,000. Therefore, a correlation between the measured cosmic-ray antiproton flux and the predicted heavier antinuclei fluxes exists. However, the heavier antinuclei formation processes are not well constrained~\cite{Gomez-Coral:2018yuk} and allow for wide span of predicted fluxes that are still compatible with the observed antiproton flux. 

{ In this section, uncertainties associated with the formation of light antinuclei, such as antideuteron and antihelium, are reviewed.} It is an important question whether antinuclei are produced in collision via freeze-out from a quark-gluon plasma (the statistical thermal model) or at a later stage via coalescence of individual antinucleons (e.g.,~\cite{Mrowczynski:2016xqm}). The results of the fixed-target experiment NA49, which measured Pb--Pb interactions at 20--158\,$A$ GeV/$c$~\cite{Anticic:2016ckv}, were in agreement by both the statistical and the coalescence models.
{At LHC energies, the analysis of ALICE data~\cite{floris, Acharya:2017fvb} suggests that the thermal model works well for Pb--Pb interactions and that $p$--Pb data~\cite{Acharya:2019xmu, Acharya:2019rys, Acharya:2019yoi} can be explained with both the coalescence model and the statistical thermal model, indicating a smooth transition. The statistical thermal model is over-predicting the deuteron-to-proton ratio for $p$--$p$ interactions. The ``light antinuclei puzzle'' is further reviewed in~\cite{Bellini:2018epz}.}

\subsubsection{Coalescence\label{s-coal}}

The fusion of an antiproton and an antineutron into an antideuteron can be described by the coalescence model, which is based on the assumption that any pair of antiproton and antineutron within a sphere of radius $p_0$ in momentum space will coalesce to produce an antinucleus~\cite{PhysRev.129.854,Sato:1981ez}. In this approach, the antideuteron spectrum is given by:
\begin{equation} \label{eq:iso_coalescence}
\gamma_{\bar d}\frac{\text d^3N_{\bar d}}{\text d p^3_{\bar d}}=\frac{4\pi}{3}p_0^3\left(\gamma_{\bar p}\frac{\text d^3N_{\bar p}}{\text d p^3_{\bar p}}\right)\left(\gamma_{\bar n}\frac{\text d^3N_{\bar n}}{\text d p^3_{\bar n}}\right),
\end{equation}
where $p_i$ and $\text dN_i/\text dp_i$ are, respectively, the momentum and the differential yield per event of particle $i$ ($\bar d=$antideuteron, $\bar p=$antiproton, $\bar n=$antineutron). The coalescence momentum $p_0$ is a critical value because it enters to the third power and directly scales the yield. It is on the order of about 100\,MeV/$c$, which is smaller than the typical scale at which the perturbative theory of quantum chromodynamics breaks down. As a result, the coalescence model is sensitive to non-perturbative effects. It is also customary to use the $B_A$ parameter to more generally describe nuclei formation with mass number $A$ in the context of heavy-ion collision measurements. It is related to $p_0$ by: 
\begin{equation}
B_A=\frac{A}{m_p^{A-1}}\left(\frac{4\pi}3p_0^3\right)^{A-1},
\end{equation}
with $m_p$ being the proton mass. As seen in Eq.~(\ref{eq:iso_coalescence}), $p_0$ can experimentally be determined from the measurement of antiproton and antideuteron spectra from the same experiment. 
Therefore, it not only describes the difference in momenta of the coalescing antinucleons, but parameterizes a number of other effects as well. It is important to note that the coalescence model is not a nuclear-physics model for the formation of light antinuclei. It should be seen as a statistical approach that is capable to reproduce the data. Therefore, a number of effects are not explicitly included in the model, but are introduced as uncertainties in the coalescence momentum, e.g., energy conservation, phase space reduction near threshold production, and  spin alignment. However, for the (anti)deuteron case energy conservation is not a primary concern due to the low binding energy. The possibility of an isospin asymmetry at energies below the LHC measured by NA49~\cite{Chvala:2003dn,Fischer:2003xh} as well as Coulomb suppression for the case of antihelium formation would be also absorbed in the coalescence momentum. Using only one constant parameter is an obvious oversimplification. Therefore, separating the actual coalescence process from other conditions is an important task for the understanding of antideuteron production. 

\begin{figure}
\centering
\includegraphics[height=0.46\linewidth]{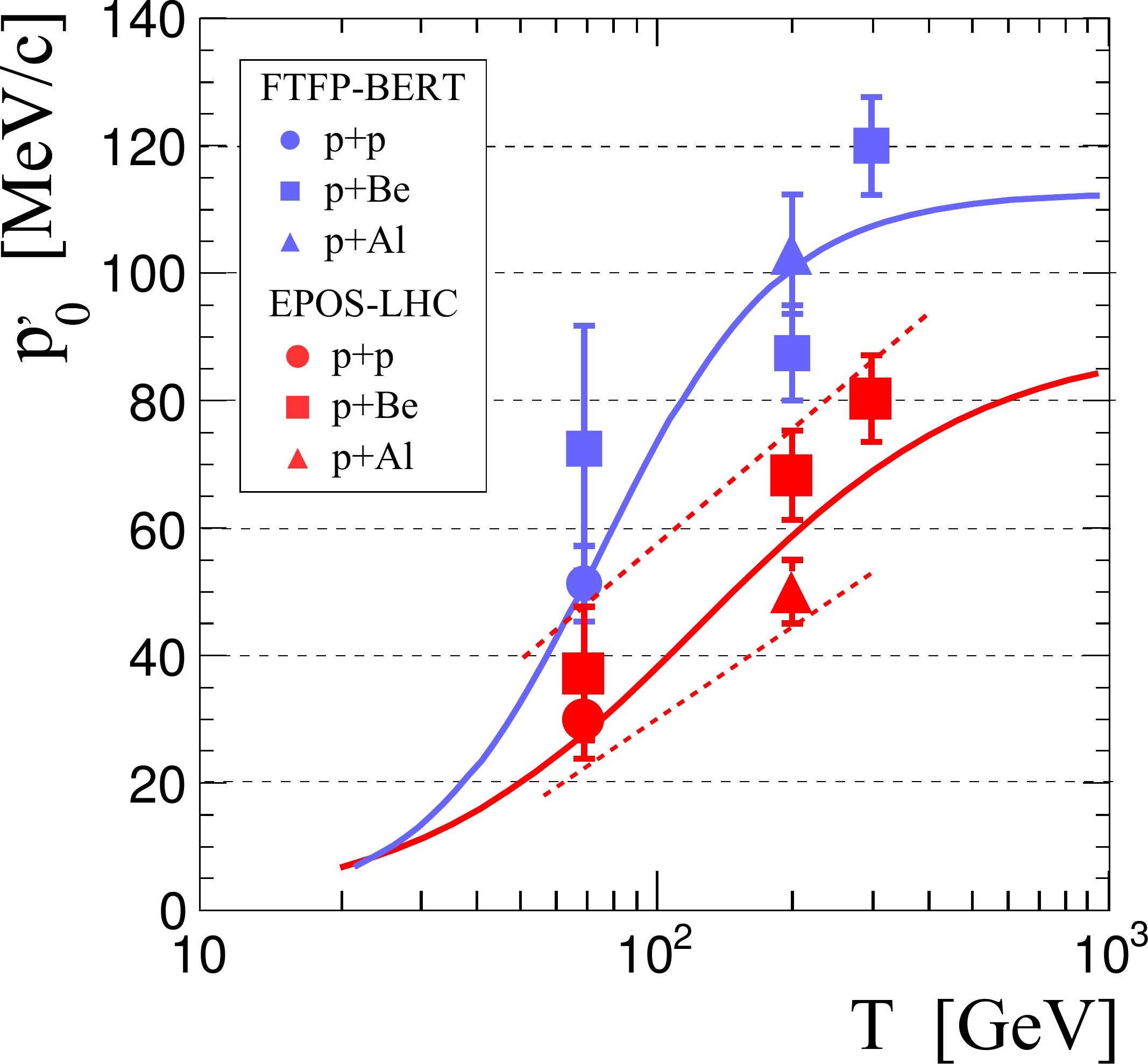}
\caption{\label{f-fig10}$p'_0$ for antideuterons as function of kinetic energy of the incoming proton for two different hadronic generators. The solid lines show empiric fits to the best-fit $p_0'$ values for the corresponding generator. The dashed red lines indicate the one-sigma uncertainty range for {\tt EPOS-LHC}~\cite{Gomez-Coral:2018yuk}.}
\vspace{-0.15in}
\end{figure}

The state-of-the-art technique is to apply the coalescence condition to $\bar{p}\bar{n}$ pairs on a per-event basis in Monte Carlo simulations. However, typical hadronic {event} generators~\cite{Pierog:2013ria,Ostapchenko:2004qz,Sjostrand:2007gs,Agostinelli2003250,2006ITNS...53..270A,Galoyan:2012bh,Ahn:2009wx} do not have the capability to produce antideuterons. Therefore, an event-by-event coalescence ``afterburner" { is implemented. Using the output of a given {event} generator, this afterburner compares the momenta of antinucleon pairs in their center-of-mass frame.} If the momentum difference is below the coalescence momentum, an antideuteron is formed. 

The formation of antideuterons is tightly coupled to the production of antiprotons. However, as discussed in Sec.~\ref{s-antipxs}, the antiproton production cross sections are not well known. 
{ A recent study~\cite{Gomez-Coral:2018yuk} compared several hadronic {event} generators, and showed that none reproduces the measured antiproton production cross sections particularly well, neither in terms of overall yield nor spectral shape.} {\tt EPOS-LHC}~\cite{Pierog:2013ria} exhibits the best agreement with the available data, but still shows an average of about two standard deviations from the measured values. These generators are not specifically tuned to reproduce { antiproton and antineutron spectra, since these make up only a small fraction of the total generated particle flux}. These inaccurate antiproton and antineutron spectra result in a shift of the derived coalescence momentum $p_0$~\cite{Gomez-Coral:2018yuk}.

In an attempt to factorize these antiproton discrepancies from the antideuteron production, the coalescence momentum in Eq.~(\ref{eq:iso_coalescence}) can be replaced with $p_0=p'_0\cdot r^\frac23$, with $r$ being a scaling factor that describes the mismatch between antiproton data and simulation for a specific generator and collision energy. 
{In this way, the redefined coalescence momentum $p'_0$ is not any longer describing the antideuteron constituent yield mismatch of the generators, but more the actual coalescence process.}
{The authors of \cite{Gomez-Coral:2018yuk} found that the coalescence momentum should be treated as a function of kinetic energy }
(Fig.~\ref{f-fig10}). It is important to note that the increase is especially steep in the region that is most relevant for cosmic rays (up to a few 100\,GeV). Historically, a constant $p_0$ value of about 80\,MeV/$c$ was assumed above the production threshold~\cite{Duperray:2005si}. 
{However, this study indicates that this assumption is inaccurate.}
Considering the spread in $p'_0$ and the antiproton mismatch correction, the total antideuteron yield has an uncertainty of more than a factor of 10. 
Furthermore, Fig.~\ref{f-fig10} also shows a general discrepancy between the models, with {\tt Geant4} being above {\tt EPOS-LHC} by about 50\%. 
The underlying uncertainties are potentially even larger: as existing antideuteron production data are very limited and the fitted coalescence momenta from $p$--$p$, $p$--Al, and $p$--Be collisions are relatively close, a function independent of the collision system was fitted. However, the $p$--Al and $p$--Be results disagree by about 2$\sigma$ and only one $p$--$p$ point in this interesting range exists. Hence, an improved understanding of $p$--$p$ interactions (50--400\,GeV/$c$) is crucial for the understanding of cosmic antinuclei fluxes. {Also the STAR experiment found an energy dependence of the $B_2$ parameter in Au--Au interactions, indicating that this parameter is not only describing the momentum difference of the coalescing (anti)nuclei constituents, but also the size of the emission of source of antibaryons~\cite{Adam:2019wnb}.}

Building on the event-by-event coalescence for antideuterons, a similar approach can be taken for antihelium formation: (a) Each possible antinucleon pair from the generator output has to fulfill the coalescence condition, (b) antinucleons are iteratively added to a multi-antinucleon state if they fulfill the coalescence condition~\cite{uhantihe3}. Using the coalescence parameter from antideuteron fits allows making direct predictions about antihelium. More dedicated studies are needed. Mixing of these two scenarios as well as other conditions are also possible. For instance, the authors of~\cite{Li:2018dxj} used a circle to envelope the relative four-momenta of the merging antinucleons, and required the diameter of this circle to be less than the coalescence momentum for antihelium-3 formation. In this approach, the coalescence momentum for antihelium needs to be determined by an additional fit to antihelium production data.

Nevertheless, { simply rescaling the simulation to match the data and ignoring the underlying reasons for the discrepancy} cannot be the final answer. As the value of $p_0$ is small, coalescence is highly sensitive to two-particle corre\-lations between the participating antinucleons. This is especially crucial for antideuteron production close to the production threshold energy, which favors an anticorrelation of antipro\-tons and antineutrons for kinematical reasons and causes phase-space suppression. The underlying mechanisms {implemented in each event generator} are very different and {one cannot expected that their} two-particle correlation models agree. There is also no a-priori reason to expect the two-particle correlations from one generator to be more reliable than from {the other}. Therefore, more {measurements} are needed to {better understand} angular correlations between antiprotons and antineutrons.

{Coalescence can also be described in a generalized phase-space coalescence approach where the light (anti)nucleus constituent wave functions are projected onto the wave function of the light antinucleus final state (e.g., ~\cite{Remler:1981du,Braun-Munzinger:2018hat}). This was recently used in an approach} in which the two-nucleon density matrix of an antideuteron is described by a two-body Wigner function that is factorized in a plane wave describing the center-of-mass motion and the internal antideuteron wave function~\cite{Kachelriess:2019taq, Kachelriess:2020uoh}. 
Wigner functions contain the full quantum mechanical information of the system. The internal deuteron wave function was approximated by Gaussian distributions that relate to the charge radius of the deuteron. This was used in combination with the nucleon production from event-by-event hadronic generators. In this case, an antideuteron is formed with a probability depending on the momentum of the merging nucleons, the RMS of the deuteron charge radius, and one free {parameter $\sigma$. The authors of \cite{Kachelriess:2019taq} find that the fit of this parameter to experimental data agrees well with its interpretation as the size of the formation region of antinuclei. This approach accounts therefore naturally for differences in the interaction region, and predicts thereby the process dependence of the coalescence parameter.}
This semi-classical model includes constraints on both momentum and space variables in a microphysical picture. 
{In addition, this model was applied to the formation of antihelium-3 and antitritium. Thus, the model describes the production of both antihelium-3 and antideuteron with a single free parameter.}

\subsubsection{Thermal Model\label{s-therm}}

In the statistical hadronization model \cite{1997hep.ph....2274B,SHM1,SHM2,SHM3,SHM4,andronic,cleymans}, light \mbox{(anti-)nuclei} as well as all other hadron species are assumed to be emitted by a source in local thermal and hadrochemical equilibrium and their abundances are fixed at chemical freeze-out, with a temperature of $T_{\text{chem}}=(156\pm4)$\,MeV~\cite{2018NuPhA.971....1A,Bellini:2018epz}:
\begin{equation}
\frac{\text dN}{\text dy} \propto \exp\left(-\frac{m}{T_{\text{chem}}}\right),
\end{equation}
with $\text dN/\text dy$ being the particle rapidity density for a particle species with mass $m$. 
Due to their large masses, the abundance of nuclei is very sensitive to $T_{\text{chem}}$.
This model provides a good description of the measured hadron yields in central nucleus--nucleus collisions and shows little dependence on energy~\cite{SHM1,2018NuPhA.971....1A}.
However, the mechanism of hadron production and the propagation of loosely-bound states through the hadron gas phase without a significant reduction in their yields are not addressed by this model. It has been proposed that such objects could be produced at the phase transition as compact colorless quark clusters with the quantum numbers identical to those of the final state hadrons. The survival of these states to high temperatures is interpreted as due to the low interaction cross section with the surrounding medium \cite{SHM1}.

The same model can be employed to describe the light-nuclei production in $p$--$p$ and $p$--Pb collisions~\cite{Acharya:2019rgc,Acharya:2020sfy, Acharya:2019rys}. 
In this case, instead of the grand-canonical implementation used for $A$--$A$ collisions, a canonical ensemble, where the baryon number ($B$), charge ($C$), and strangeness ($S$) must be exactly conserved, has to be used. The model qualitatively reproduces the trend observed in data with $T_{\text{chem}} = 155$\,MeV, $B = Q = S = 0$ and two different values of the correlation volume $V_c$~\cite{Acharya:2020sfy}. 
This could suggest that for small collision systems the light (anti)nuclei production could 
canonically be suppressed and that a canonical correlation volume might exist. However, the correlation volume used in these calculation is $V_c=\text{d}V/\text{d}y = (2.4 \pm 0.2)\ \text{fm}^3\cdot\text{d}N_{ch}/\text{d}\eta$, with the pseudorapidity density $\text{d}N_{ch}/\text{d}\eta=6.01\pm0.01$ for charged particles at $\sqrt{s}=7$\,TeV~\cite{ALICEmult7}. This $V_c$ is at least 2.5 larger that the one obtained by femtoscopy measurements~\cite{Acharya:2018gyz, Acharya:2019bsa}, where a source radius of $r_0 = (1.125 \pm 0.018\ (\text{stat.})^{+0.058}_{-0.035}\  (\text{syst.}))$\,fm was measured. This could indicate that the volume used in the thermal model calculation might be overestimated and that the thermal model calculations have to be checked against the  precise radii measurements available in $p$--$p$ and $p$--Pb collision~\cite{Acharya:2019sms, Acharya:2019yvb}.

A related model for the antideuteron production, based on Hanbury-Brown-Twiss (HBT) interferometry, was introduced in~\cite{Scheibl:1998tk} for heavy-ion collisions and recently more generalized in~\cite{Blum:2019suo}. This model takes into account the size of the emitting volume. The larger the distance between the antiprotons and antineutrons created in the collision, the less likely it is that they coalesce. The source can be parameterized as rapidly expanding under radial flow. The coalescence process is then governed by the same correlation volume, which can be extracted from HBT interferometry (e.g,~\cite{BOGGILD1993510}). The source radius enters in the quantum-mechanical correction factor that accounts for the size of the object being produced. The advantage is that this model does not contain free parameters and predicts different $B_2$ values for different source sizes ($p$--$p$: 0.5-1.2\,fm, $p$-$A$: 1.25-1.58\,fm),  which describes the data well. This was used to make predictions for $B_3$ in~\cite{Blum:2017qnn} to explain the AMS-02 antihelium candidate events by using the most optimistic yield value. A joint coalescence-HBT analysis in transverse moment and centrality bins may further reduce uncertainties.

As discussed in Sec.~\ref{s-coal}, typical hadronic generators do not produce light nuclei. A modified approach to the standard {\tt EPOS} technique is currently in the early stages. {{\tt EPOS} incorporates two types of hadronization.}  The first one is the standard string hadronization in the corona of low density without the creation of light antinuclei. The second one is the hadronization in the core of high density, which can be seen as a collective or thermal process and produce antinuclei directly without producing antinucleons first. Initial studies show a good description of light antinuclei production in heavy ion collisions. An advantage of this model is that the collective hadronization parameters are constrained by other types of particles that then make predictions about antinuclei formation.

\subsection{Cosmic-ray Propagation\label{s-prop}}

\subsubsection{Galactic Diffusion}

Important constraints for the antinuclei flux from dark matter annihilations are coming from the values of the diffusion coefficient, its rigidity dependence, and the Galactic halo size, which directly scales the observable flux~\cite{Donato:2003xg}. Fits of cosmic-ray nuclei data for secondary-to-primary ratios (e.g., Li/C, Li/O, Be/C, Be/O, B/C, B/O) are important to constrain propagation models~\cite{1990acr..book.....B,2007ARNPS..57..285S,2017ApJ...840..115B,Reinert:2017aga,2018ApJ...854...94B,2018ApJ...858...61B,Boudaud:2019efq,2019arXiv191103108B}. However, this approach is hampered by uncertainties in the production  cross sections at the level of 10--20\%~\cite{0067-0049-144-1-153,2010A&A...516A..67M,2015A&A...580A...9G,Tomassetti:2017hbe,Genolini:2018ekk,Evoli:2019wwu}. These uncertainties propagate directly into predictions for the fluxes. For instance, if the effective column density is derived from the measured B/C ratio with 20\% too low cross section for boron production, the predicted secondary antimatter fluxes will be 20\% too high.
These secondary-to-primary flux ratio fits are also somewhat degenerate in the ratio of the diffusion coefficient normalization to the halo size, which can be broken if radioactive secondaries like Beryllium-10 are used~\cite{2007ARNPS..57..285S}.

In addition to the standard Galactic transport model~\cite{1990acr..book.....B,2007ARNPS..57..285S}, a number of alternative or expanded propagation models have been discussed. For instance, secondary particles could be accelerated in supernova remnant shocks. However, this is in tension with the spectra of secondaries, which are steeper than primaries in the whole energy range. Furthermore, non-linearities or local variations in the propagation parameters can influence the cosmic-ray spectra without adding ad-hoc breaks into the injection spectra~\cite{2012ApJ...752...68V,2012PhRvL.109f1101B,2018PhRvL.121b1102E,2019arXiv191103108B}. Some of the studies are restricted by the precision and availability of production cross section measurements for interactions of relevant primary cosmic-ray species with the interstellar gas~\cite{Genolini:2018ekk}. As pointed out above, improvement in the accuracy of the  production cross sections is the key to for the understanding of cosmic-ray antinuclei as well. In the future, deuterons, helium-3, and sub-iron nuclei can be used to study propagation effects on different spatial scales~\cite{Trotta:2010mx}.

\subsubsection{Solar Modulation}

\begin{figure}
  \begin{center}
    \includegraphics[width=0.8\linewidth]{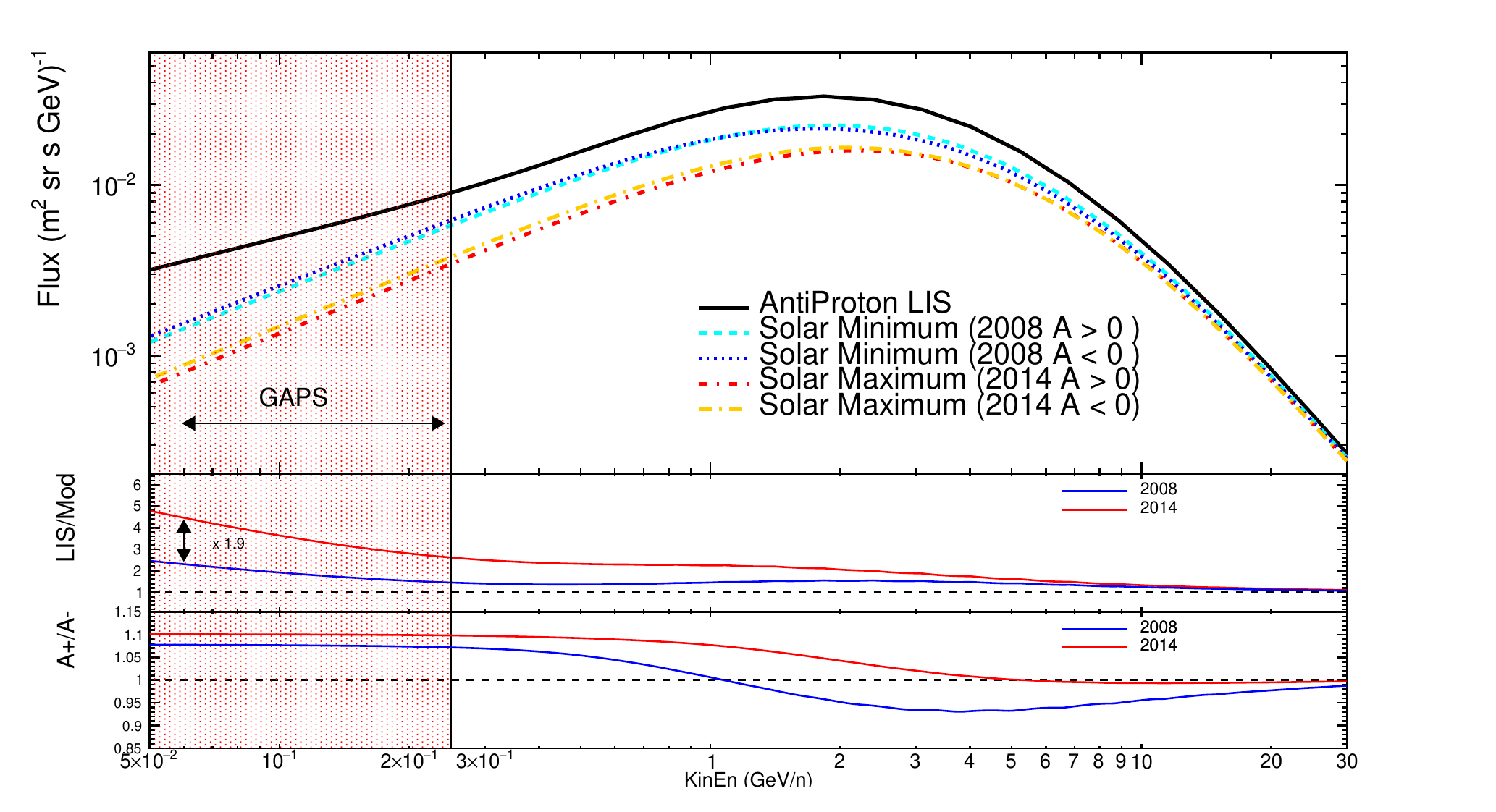}
    \end{center}
\caption{\label{f-fig11}\textbf{Top panel:} antiproton flux for solar minimum and maximum from a 3D numerical model for cosmic-ray propagation through the Heliosphere. The solar minimum and solar maximum spectra has been evaluated for positive and negative polarity of the HMF ($A>0$ and $A<0$) to show the effect of the charge sign solar modulation. \textbf{Middle panel:} ratio between the LIS and the modulated spectra for the period of maximum and minimum solar activity ($A<0$). \textbf{Bottom panel:} ratio between the modulated spectra for opposite polarity of the HMF for the period of minimum and maximum solar activity.}
\end{figure}

Most of cosmic ray measurements are made deep inside the heliosphere, at or near the Earth, and thus are affected by the heliospheric (a.k.a. solar) modulation~\cite{2018ApJ...854L...2M,PhysRevLett.121.051102}. The latter affects the spectra of cosmic-ray species below 50\,GV, with the modulation level depending on the solar activity, configuration and polarity of the solar magnetic field~\cite{1965P&SS...13....9P}. Many aspects of the solar modulation were understood a while back, but the absence of accurate heliospheric data, the local interstellar measurements, and computer resources resulted in simplifications and extrapolations. For example, the well-known force-field approximation~\cite{gleeson-1967}, which accounts for adiabatic energy losses and has only one parameter, the so-called modulation potential, was extensively used in the past. Even though it can fit the observed data with one free parameter quite well, it has no predictive power and cannot account for charge-sign effects. 

Cosmic-ray propagation in the heliosphere was first studied by Parker~\cite{1965P&SS...13....9P}, who formulated the transport equation, also referred to as Parker equation. It includes diffusion, adiabatic energy changes, convection, and drift effects. High precision data from recent space missions, ACE, PAMELA, AMS-02, together with observations from Ulysses spacecraft outside the ecliptic plane and from Voyager 1 and 2, which probe the outer reaches of the heliosphere, enabled more sophisticated 3D modeling of heliospheric propagation~\cite{Potgieter:2013bra,DiFelice:2016oec,2017ApJ...840..115B,2018ApJ...854L...2M,2018ApJ...854...94B,2018ApJ...858...61B,Aslam:2018kpi,2019arXiv191103108B,2019AdSpR..64.2459B}. Their data taken at different levels of solar activity and polarities of the solar magnetic field allow a derivation of the local interstellar spectra~\cite{2017ApJ...840..115B,2018ApJ...854...94B,2018ApJ...858...61B,2019arXiv191103108B,2019AdSpR..64.2459B} of cosmic-ray species and significantly more reliable calculation of the modulated spectra even during the periods of active Sun.

Fig.~\ref{f-fig11} (top panel) shows the antiproton spectrum predicted with one of these solar modulation models~\cite{Aslam:2018kpi} for the 2008 solar minimum and the 2014 solar maximum. The model takes as input the antiproton Local Interstellar Spectrum (LIS) and calculates the intensity at Earth for the corresponding values of the Heliospheric Magnetic Field (HMF), solar wind velocity, tilt angle, and all the other relevant Heliospheric parameters. During a period of solar minimum as in 2008, below 200\,MeV the model predicts (Fig.~\ref{f-fig11}, middle panel) an antiproton flux a factor 2 higher than a period of solar maximum as in 2014. The model also predicts a difference of around 10\% for antiprotons calculated during the same period of solar activity, but with opposite HMF polarity, as can be seen in Fig.~\ref{f-fig11} (bottom panel). In terms of solar modulation, this is equivalent to invert the charge sign of the particle with the same HMF polarity. This means that a solar modulation model, which does not include the charge sign dependence, introduces systematic uncertainties of about 10\%. Therefore, a precise study of dark matter signatures in the antiparticle spectra requires a  proper treatment of propagation in the heliosphere via advance numerical models that include the latest theoretical developments concerning turbulence and diffusion theory.

\subsection{Prospects for Ground-based Measurement for Reducing Uncertainties}

{ Improved antinuclei production cross section measurements using ground-based experiments are critical to reducing the antinuclei production and propagation uncertainties discussed above.} Antihelium-3 { and antideuteron} production measurements at LHC energies with ALICE are already very useful to test formation models. However, using different collision systems with energies closer to the production threshold of light antinuclei is necessary to understand their production in the Galaxy. The following summarizes ongoing and potential future efforts in this direction.

The fixed-target experiment NA61/SHINE (SPS Heavy Ion and Neutrino Experiment) at the Super Proton Synchrotron (SPS) at CERN is a hadron spectrometer capable of studying collisions of hadrons with different targets over a wide incident beam momentum~\cite{na61}. The detector consists of different subdetectors for the particle identification. NA61/SHINE already recorded $p$--$p$ interactions with beam momenta from 13 to 400\,GeV/$c$, and also collected data for other hadron interactions, including $p$--C, $\pi^+$--C, Ar--Sc, $p$--Pb, Be--Be, Xe--La, Pb--Pb at different energies. NA49 and NA61/SHINE have published several relevant data~\cite{na49Antiprotons,Aduszkiewicz:2017sei} that are used for tuning cosmic-ray formation and propagation models (Secs.~\ref{s-antipxs} and \ref{s-prop}). The analysis of large $p$--$p$ data sets at 158\,GeV/$c$ and 400\,GeV/$c$ are of particular interest. The measurement of light nuclei in various $A$--$A$ data sets can be used to study the production of light ions at the threshold. These measurements will complement the NA49~\cite{dbarna49,Anticic:2016ckv} and ALICE results and allow to test the coalescence and thermal models in a different regime. Also a first pilot run of carbon fragmentation measurement was conducted by the end of 2018 and demonstrated that the measurements are possible~\cite{2019arXiv190907136U}. Extended data taking with NA61/SHINE will be possible in the future.

Furthermore, it is proposed to use the COMPASS++/AMBER experiment at SPS to perform measurements with protons between 50 and 280\,GeV/$c$ on fixed liquid hydrogen and helium targets~\cite{Denisov:2018unj}. The experiment is a magnetic spectrometer consisting of a number of subdetectors for the particle identification (including ring image Cherenkov, electromagnetic and hadronic calorimeters, gas electron multipliers). The effective data rate of protons with a 40\,cm long liquid hydrogen target is estimated to be in the range of 30\,kHz. A full simulation study was carried out to estimate the acceptance and efficiency. For 20\,bins in momentum from 10 to 50\,GeV/$c$ and 20\,bins in transverse momentum with 75\% beam purity at $5\cdot10^5$ protons-per-second beam intensity, it is expected that the statistical error is on the percent level for most of the differential cross section bins within several hours of beam time for each energy. The program could start after the LHC long shutdown in 2021. Preliminary measurements are expected to be performed in 2021 and 2022.

Also ALICE will continue studies of light antinuclei production in the next years. 
ALICE is a LHC experiment that makes use of specific energy loss, time-of-flight, transition radiation, Cherenkov radiation, and calorimetric measurements for the particle identification~\cite{Abelev:2014ffa}. As discussed in Secs.~\ref{s-antipxs} and \ref{s-antinuclprod}, ALICE data are already actively used for constraining antinuclei production models. Future results will also include interaction cross section measurements of the produced antinuclei with the detector materials. Only very little data exist in this regard and it will be important to inform the analysis and interpretation of cosmic-ray data.

In addition to the $p$--He measurements of LHCb already discussed in Sec.~\ref{s-antipxs}, the experiment also aims at antideuteron production cross section measurements in $p$--$p$ interactions. LHCb is unique in the sense that it measures in the very forward direction ($2<\eta<5$). For particle identification it uses a vertex locator around the collision point, ring imaging Cherenkov detectors, electronic and hadronic calorimeters, tracking stations, and muon stations. Antideuterons can be measured at LHCb in prompt production in $p$--$p$ collisions, in decays of heavy-hadrons, and in fixed target collisions.

\section{Prospects for Future Experiments\label{s-future}}

In the coming decades, the development of new experimental techniques holds the promise of improved sensitivity to cosmic-ray antinuclei.
This section reviews ideas for future detector concepts.

\subsection{The GRAMS Experiment}

\begin{figure}
  \begin{center}
    \includegraphics[width=0.5\linewidth]{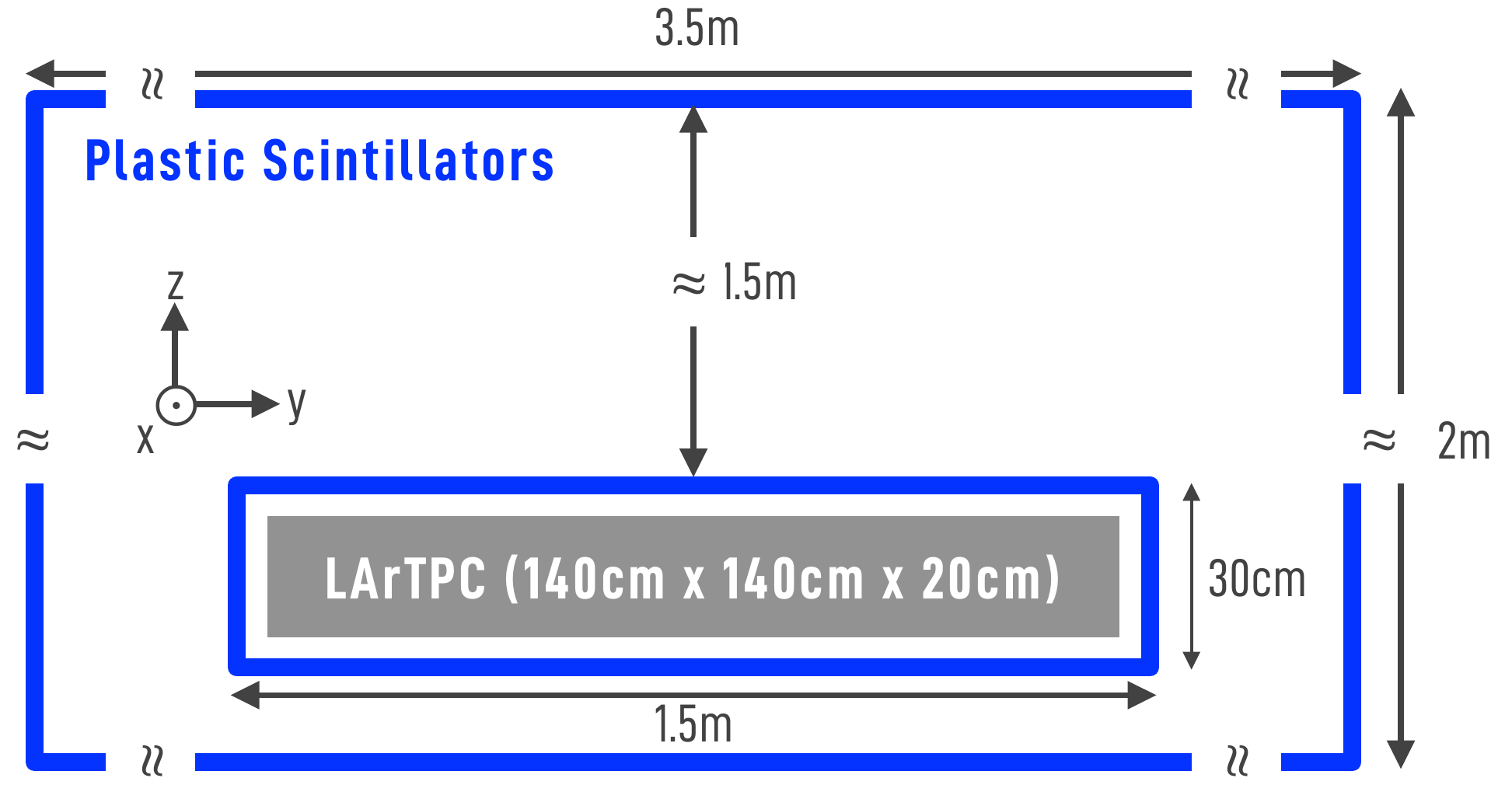}
    \end{center}
\caption{\label{f-fig12} The GRAMS detector consists of a segmented liquid-argon time projection chamber surrounded by plastic scintillators. Reprinted with permission from~\cite{Aramaki:2019bpi}.}
\end{figure}

The GRAMS (Gamma-Ray and AntiMatter Survey~\cite{Aramaki:2019bpi}) experiment is a novel instrument designed to simultaneously target both astrophysical gamma-rays with MeV energies and antimatter signatures of DM. The GRAMS instrument, shown in Fig.~\ref{f-fig12}, consists of a liquid-argon time projection chamber (LArTPC) surrounded by plastic scintillators. The LArTPC is segmented into cells to localize the signal, which is an advanced approach to minimize coincident background events in the large-scale LArTPC detector. 

The GRAMS concept potentially allows for a larger instrument, since argon is naturally abundant and low cost, than current experiments that rely on semiconductor or scintillation detectors. GRAMS is proposed to begin as a balloon-based experiment as a step forward to a satellite mission. A reasonable development timescale for GRAMS would put its first balloon flight just beyond the completion of the three-light GAPS mission and the future COSI-X mission.

A single GRAMS LDB flight could provide an order of magnitude improved sensitivity to astrophysical MeV gamma-rays, which have not yet been well-explored.
This will allow study of the formation, evolution, and acceleration mechanisms of astrophysical objects, 
in particular multi-messenger signals associated with gravitational waves from merging massive objects
and nuclear line signals of Galactic chemical evolution and supernovae.

In addition, GRAMS has been developed to become a next-generation search for antimatter signatures of DM. The detection concept is similar to that of GAPS, relying on exotic atom capture and decay. As the LArTPC detector can provide an excellent 3-D particle tracking capability with nearly no dead volume inside the detector, the detection efficiency can be significantly improved. Furthermore, the LArTPC system, which is cost-effective and easy to replace, allows for quick refurbishment and turnaround of balloon flights.

\subsection{The ADHD Experiment}

The goal of the AntiDeuteron Helium Detector (ADHD) is to use the distinctive signature of delayed annihilation of antinuclei in helium to identify cosmic antimatter species. 
The typical lifetime for stopped antideuterons in matter is on the order of picoseconds, similar to that of stopped antiprotons. However, the existence of long-lived (on the order of microseconds) metastable states for stopped antiprotons in helium targets has been measured~\cite{PhysRevLett.67.1246}. These metastable states in helium have also been measured for other heavy negative particles, such as pions and kaons~\cite{1992PhRvA..45.6202N,1989PhRvL..63.1590Y}. The theoretical description of this effect predicts that the lifetimes of these metastable state increase quadratically with the reduced mass of the system, i.e., a larger delay of the annihilation signature is expected for antideuteron capture in helium than for antiproton capture~\cite{osti_4029624,PhysRevLett.23.63,PhysRev.188.187,PhysRevA.6.2488,PhysRevA.51.2870}.

\begin{figure}
  \begin{center}
    \includegraphics[width=0.5\linewidth]{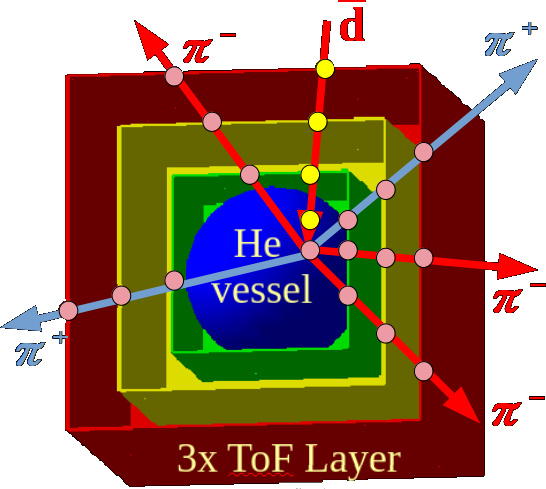}
    \end{center}
\caption{\label{f-fig13} The ADHD instrument is made of a 400\,bar helium-gas calorimeter (HeCal) surrounded by a time-of-flight system with three scintillator layers. Antideuterons are detected as a single in-going prompt track (yellow points) followed by several out-going delayed pions (light-red points).}
\end{figure}

A possible layout for ADHD is shown in Fig.~\ref{f-fig13}.
The inner portion contains the helium calorimeter (HeCal), which consists of an approximately 90-cm diameter spherical thermoplastic vessel filled with 20\,kg of scintillating helium (300 liters at 400\,bar pressure). The vessel wall thickness of about 3\,cm (mass about 100\,kg) ensures a burst pressure larger than 800\,bar; a similar system is already considered for helium transportation in space or for H$_2$ tanks~\cite{h_tank}. Helium gas is a fast UV scintillator, having a light yield similar to other fast plastic or liquid scintillators and capable of providing nanosecond timing performance~\cite{doi:10.1063/1.3665333}. The HeCal is surrounded by a time-of-flight system consisting of three layers of 4\,mm-thick plastic scintillator bars, which provides velocity and charge measurements via ionization energy loss. It is assumed that with current technology such a system can deliver velocity resolution of 5\% and energy-loss resolution of 10\%, implying a timing resolution on the order of 100\,ps and position resolution of the order of a few centimeters. 

The expected acceptance is about 0.2\,m$^2$sr in the energy region 60--150\,MeV/$n$ for antideuterons and 100--300\,MeV for antiprotons.
Considering energy loss in the TOF and vessel, a minimum kinetic energy of approximately 60\,MeV/$n$ is necessary for antideuterons to reach the helium target. On the other hand, antideuterons with kinetic energy larger than 150\,MeV/$n$ would typically cross the active helium region without stopping inside. 
Combining information on the velocity and energy depositions measured by the TOF, the prompt and delayed energy measured by the HeCal, and the reconstructed event topology, it is possible to identify a single antideuterons over $10^3$ background antiprotons. 

Beam tests using a helium scintillator prototype (200\,bar Arktis Radiation Detectors B670~\cite{ARKTIS}) are ongoing to validate the expected ADHD performance.

\subsection{The AMS-100 Experiment}

\begin{figure}
  \begin{center}
    \includegraphics[width=0.8\linewidth]{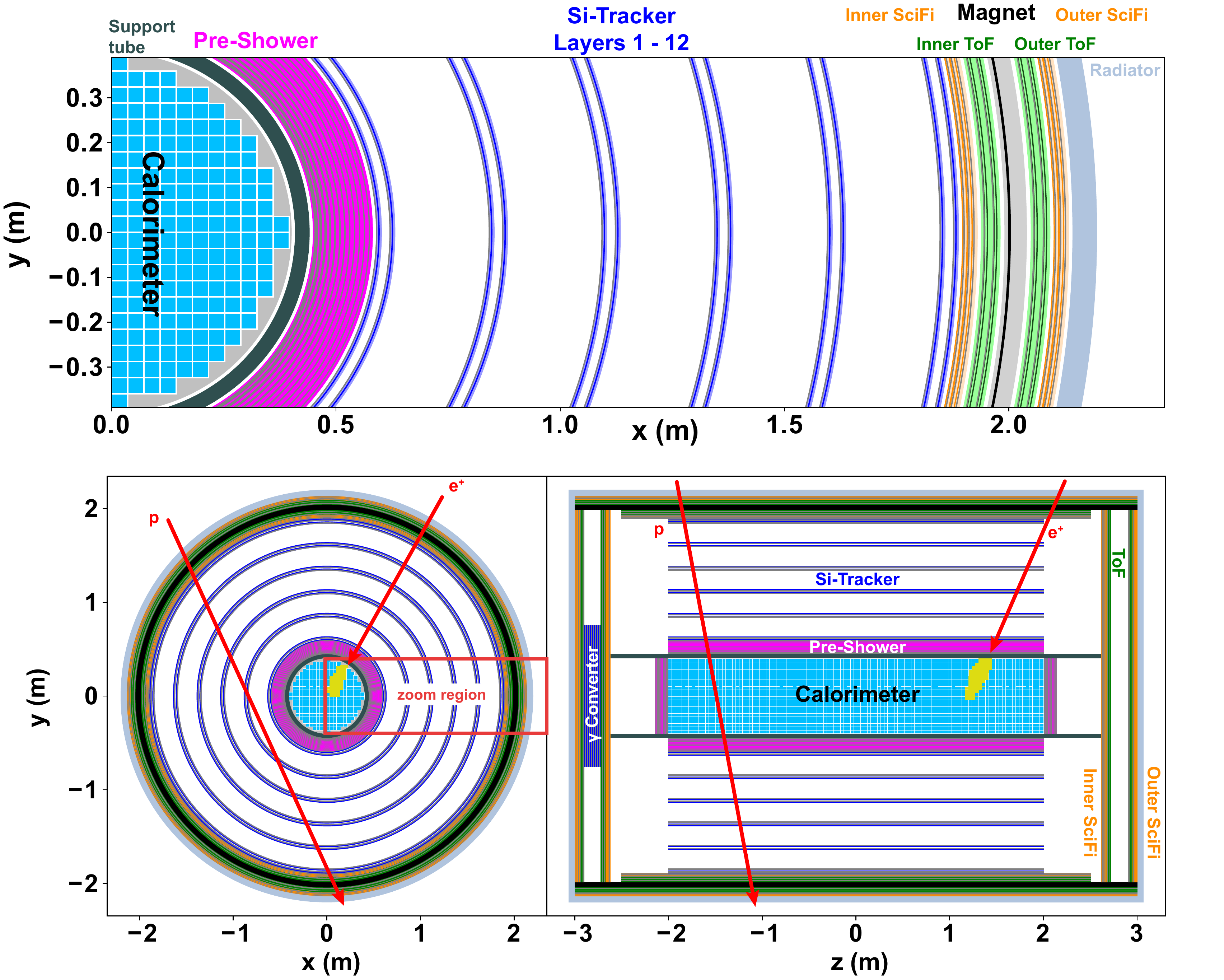}
    \end{center}
\caption{\label{f-fig14} The AMS-100 instrument consists of a central calorimeter surrounded by a silicon-strip and scintillating-fiber tracker, all within a high-temperature superconducting solenoid magnet. Reprinted with permission from~\cite{2019NIMPA.94462561S}.}
\end{figure}

The Next-Generation Spectrometer (AMS-100), shown in Fig.~\ref{f-fig14}, is an idea for a space-based international platform for precision particle astrophysics and fundamental physics that will vastly improve on existing measurements of cosmic rays and $\upgamma$-rays~\cite{2019NIMPA.94462561S}. Achieving this will require overcoming a number of key technical challenges, though in many cases technology already exists that can be adapted to provide the needed performance. 

The key component of the instrument is a thin, large-volume, high-temperature superconducting solenoid magnet, which provides a uniform 1\,T field within the tracking volume. When instrumented with proven silicon-strip and scintillating fiber technologies, the spectrometer can achieve a maximum detectable rigidity (defined by $\sigma_R/R = 1$) of 100\,TV, with an effective acceptance of 100 m$^2$sr. A deep (70\,$X_0$/4\,$\lambda_I$) central calorimeter provides energy measurements and particle identification.  

A Magnetic Spectrometer with a geometrical acceptance of 100\,m$^2$sr (AMS-100) will probe the antideuteron spectrum in the range 0.1-8\,GeV/$n$ with $10^5$ higher sensitivity than current experiments and  greatly expand the sensitivity to heavy cosmic antimatter ($Z\leq-2$). The magnet and detector systems will be designed with no consumables, allowing for an extended 10-year payload lifetime at its thermally-favorable orbital location at Sun-Earth Lagrange Point 2 (L2).

\section{Conclusions and Outlook for the Coming Decade}

The field of cosmic-ray antinuclei measurements is currently undergoing a transformation, with experiments offering breakthroughs in sensitivity to viable DM models.
In the coming decade, AMS-02 will continue accumulating the large statistics and systematic understanding necessary for it to probe rare antinuclei signals, and GAPS, which is the first experiment optimized specifically for low-energy cosmic antinuclei, will begin several Antarctic balloon campaigns.

The high-statistics antiproton spectral measurements from AMS-02 have already provided leading constraints on thermal WIMP DM and revealed a possible excess in its low-energy spectrum. 
This excess is consistent with a range of DM scenarios, such as a canonical WIMP with a mass in the range of 40--130\,GeV, and with constraints from direct detection experiments, though its significance varies depending on the interpretation of systematic uncertainties, in particular careful treatment of any energy correlations.
In its first flight, GAPS will measure a precision antiproton spectrum in a low-energy ($E<0.25$\,GeV/$n$) region currently inaccessible to any experiment, opening sensitivity to new DM models. 
These antiproton results provide valuable input for flux predictions for heavier antinuclei species. 

The potential detection of candidate antihelium-3 and antihelium-4 events by AMS-02 is still tentative and the analysis is still ongoing.
These studies will also clearly benefit from extended data taking. 
Given the transformative nature of such a discovery, verification with a complementary experimental technique such as GAPS is important.
If taken seriously, these events motivate optimistic coalescence scenarios (which are in tension with current ALICE and heavy-ion measurements) and specifically-tuned DM models that can boost any expected antihelium signal to be within the detection range of AMS-02 without overproducing antiprotons or antideuterons.
However, even the most optimistic coalescence scenarios cannot account for the detection of a single cosmic antihelium-4 nucleus from either astrophysical or DM origins.
Instead, these events could be due to local antimatter over-densities resulting from anisotropies in Big-Bang nucleosynthesis, in particular local antistars. 

Predictions for low-energy antideuterons, which have been recognized for over a decade as  a potential ``smoking gun'' signature of DM annihilation or decay, are being revisited in light of these recent antiproton and antihelium results. 
Also the AMS-02 antideuteron analysis is ongoing and will provide sensitivity to the relevant flux ranges of antideuterons, whose secondary production is strongly kinematically suppressed at low energies. Within the next few years GAPS will deeply probe the low-energy antideuteron flux at lower kinetic energies than AMS-02.
Investigating the associated production of antideuterons is the most direct option to cross-check the possible excess of antiprotons, and antideuterons also provide a crucial independent channel to constrain models of antihelium production. 
In particular, any increase in coalescence probability, which is necessary to explain the antihelium-3 signal via astrophysical or DM sources, would necessarily also boost the antideuteron flux. 
Antideuterons can also provide sensitivity to DM signatures that would not be visible in antiprotons, due to the abundant antiproton secondary background, or in antihelium, due to suppression of heavier antinuclei formation as a result of coalescence. 
A range of simple DM models that are invisible to collider and direct detection methods, such as hidden sector theories with a dark photon, will be testable by antideuteron signatures. 
Perhaps more important than any particular motivating model, the history of cosmic-ray experiments emphasizes that {it is crucial to} 
remain open to serendipitous, unexpected results when first probing a new cosmic signal {or dramatically improve sensitivity of the instrument}.

Given the increasing experimental sensitivity to cosmic antinuclei, new studies are pushing to constrain the relevant systematic uncertainties from production cross sections, antinuclei coalescence, and propagation {of cosmic-ray species in the interstellar and heliospheric} enviornments. 
The relative theoretical uncertainties of antideuteron formation and {their} propagation are currently both on the order of 10. In the light of the AMS-02 antihelium candidate events, it is especially important to reduce the antinuclei formation uncertainties because typically
every process capable of producing heavier antinuclei should produce antiprotons
in much higher quantity, with each additional antinucleon reducing the yield by about a
factor of 1,000. Improved results from ground-based measurements, like NA61/SHINE, COMPASS++/AMBER, ALICE, and LHCb, will help to break degeneracies between different models trying to explain the antihelium events.

Hence, we are at the cusp of an exciting time for experimental searches for cosmic-ray antinuclei. 
To fully exploit the potential of these measurements,  the lifetimes of current missions must be supported and, in parallel, future mission technologies and concepts must be developed. Within the AMS-02 program, antideuteron and antihelium searches are among the measurements that will most significantly benefit from extended data taking on the ISS. 
The baseline sensitivity of the current GAPS design is foreseen for three 35-day long-duration balloon flights, and an upgraded payload design compatible with either rapid relaunch or approximately 100-day ultra-long duration balloon capabilities could provide over an order-of-magnitude improvement in sensitivity. 
Future mission concepts, such as GRAMS and ADHD, further push the discovery space for antideuterons and antihelium. 
Successor experiments such as AMS-100 could move from discovery to measuring spectra with high statistics. 
This is only possible with extended data taking and larger payloads, e.g., on the Moon's surface or at one of the Lagrange points.

\section*{Acknowledgments}
This work is supported in the U.S. by NASA APRA grants (NNX17AB44G, NNX17AB45G, NNX17AB46G, and NNX17AB47G). 
R.A. Ong received support from the UCLA Division of Physical Sciences.
P. von Doetinchem received support from the National Science Foundation under award PHY-1551980.
H. Fuke is supported by JSPS KAKENHI grants (JP17H01136 and JP19H05198) and Mitsubishi Foundation research grant 2019-10038.
N. Fornengo, F. Donato, and M. Korsmeier are supported by the ``Departments of Excellence'' 2018-2022 grant awarded by the Italian Ministry of Education, University and Research (\textsc{miur}) L.\ 232/2016, research grant TAsP (Theoretical Astroparticle Physics) funded by INFN, research grant `The Dark Universe: A Synergic Multimessenger Approach' No.\ 2017X7X85K funded by MIUR. N. Fornengo is further supported by the research grant ``The Anisotropic Dark Universe'' No.\ CSTO161409, funded by Compagnia di Sanpaolo and University of Turin.
M. Kachelriess' and J. Tjemsland's work was partially supported by the Research Council of Norway (NFR).
I.V. Moskalenko acknowledges a partial support from NASA APRA grant NNX17AB48G.
M. Naskret's work was supported by the Polish National Science Centre grant 2017/27/N/ST2/01009.
F. Rogers is supported through the National Science Foundation Graduate Research Fellowship under Grant No. 1122374.
M.W. Winkler acknowledges support by the Vetenskapsradet (Swedish Research Council) through contract No. 638-2013-8993.

\providecommand{\href}[2]{#2}\begingroup\raggedright\endgroup

\end{document}